\documentclass[11pt,a4paper]{scrartcl}
\pdfoutput=1
\usepackage[latin1]{inputenc}							
\usepackage{amsmath,amsfonts,amssymb}	
\usepackage{multicol}
\usepackage{graphicx}
\usepackage{natbib}
\usepackage{subfigure}
\usepackage{color}
\usepackage{amssymb}
\usepackage[colorlinks=true,citecolor=blue]{hyperref}

\makeatletter							
\def\printtitle{
    {\centering \huge \normalfont \textbf{\@title}\par}}		
\makeatother							

\title{Inertial and dimensional effects on the instability of a thin film}
\makeatletter							
\def\printauthor{
    {\large \@author}}				
\makeatother							
\author{
	Alejandro G. Gonz\'alez\\
	{\small Instituto de F\'{\i}sica Arroyo Seco,\\
	CIFICEN-CONICET,\\
	Universidad Nacional del Centro
	de la Provincia de Buenos Aires,\\ Pinto 399, Tandil, Argentina\\
	\texttt{}}\vspace{10pt} \\
	Javier A. Diez \\
	{\small Instituto de F\'{\i}sica Arroyo Seco,\\
	CIFICEN-CONICET,\\
	Universidad Nacional del Centro
	de la Provincia de Buenos Aires,\\ Pinto 399, Tandil, Argentina\\
	\texttt{}}\vspace{10pt} \\
	Mathieu Sellier \\
	{\small Mechanical Engineering Department,\\ University of Canterbury, \\
		Christchurch 8140, New Zealand\\
		\texttt{}}
	}

\usepackage{fancyhdr}												
\usepackage{lastpage}

\setkomafont{sectioning}{\rmfamily\bfseries\boldmath}

\usepackage[runin]{abstract}			
\abslabeldelim{\quad}						
\begin{document}

\printtitle 
	\vspace{0.5cm}
\begin{minipage}{0.35\linewidth}
	\begin{flushright}
		\printauthor
	\end{flushright}
\end{minipage} \hspace{0pt}
\begin{minipage}{0.02\linewidth}	
	\rule{3pt}{275pt}
\end{minipage} \hspace{0pt}
\begin{minipage}{0.63\linewidth}
\vspace{5pt}
\begin{abstract}
We consider here the effects of inertia on the instability of a flat liquid film 
under the effects of capillary and intermolecular forces (van der Waals
interaction). Firstly, we perform the linear stability analysis within the long
wave approximation, which shows that the inclusion of inertia does not produce
new regions of instability other than the one previously known from the usual
lubrication case. The wavelength, $\lambda_m$, corresponding to he maximum
growth, $\omega_m$, and the critical (marginal) wavelength do not change at all.
The most affected feature of the instability under an increase of the Laplace
number is the noticeable decrease of the growth rates of the unstable modes. In
order to put in evidence the effects of the bidimensional aspects of the flow
(neglected in the long wave approximation), we also calculate the dispersion
relation of the instability from the linearized version of the complete
Navier-Stokes (N--S) equation. Unlike the long wave approximation, the
bidimensional model shows that $\lambda_m$ can vary significantly with inertia
when the aspect ratio of the film is not sufficiently small. We also perform
numerical simulations of the nonlinear N--S equations and analyze to which
extent the linear predictions can be applied depending on both the amount of
inertia involved and the aspect ratio of the film.
\end{abstract}
\end{minipage}
\vspace{10pt}

\section{Introduction}
The stability of thin films on substrates has been for a long time a basic
subject of research, not only because of the numerous technological
applications, including coatings, adhesives, lubricants, and dielectric layers,
but also because of their fundamental
interest~\citep{craster_09,eggers_rmp97,oron_rmp97}. The dewetting of thin
liquid films is the process of destabilization of such films which leads to the
formation of drops. It is generally observed when the supported liquid film is
placed on a substrate under partial wetting conditions, and subject to
destabilizing intermolecular forces. For a homogeneous isotropic liquid on a
uniform solid substrate, two main dewetting processes are known: (i) the
nucleation of holes at defects or dust particles, and (ii) the amplification of
perturbations at the free surface (e.g., capillary waves) under the
destabilizing effect of long-range intermolecular forces in the so-called
spinodal dewetting~\citep{thiele_epje03,thiele_prl01,thiele_prl98}. Although the
distinction between these two dewetting processes is well established in the
literature, there is still a lot of debate about which of these mechanisms is
actually observed in a given experiment.

In this context, lubrication approximations to the full Navier-Stokes equations
have shown to be extremely useful for investigating the dynamics and instability
of thin liquid films on substrates, including the motion and instabilities of
their contact lines~\citep{oron_rmp97}. The theoretical treatment of the coating
problem is greatly simplified if the film is so thin that the lubrication
approximation can be employed. When this modeling is valid, it is possible to
determine the velocity field of the liquid as a function of the film thickness,
and the problem reduces to the solution of a nonlinear evolution equation for
the thickness profile of the film. To leading order, at low speeds, the dynamics
is controlled by a balance among capillarity, viscosity, and intermolecular
forces, without inertia playing a role. This approach has achieved considerable
success in dealing with the solution of this class of
problems~\citep{colinet_epj2007}. 

However, in some applications such as the dewetting of nano-scale thin metallic
films on hydrophobic substrates, the effects related to inertia and the
shortcomings of the lubrication approximation assumptions (requiring small
slopes and consequently small contact angles) appear to have a crucial influence
on the dynamics and morphology of the film~\citep{gonzalez_lang2013}.
Experimental studies of unstable thin films coating solids have shown
significant differences in the patterns that develop when fluid instabilities
lead to the formation of growing \emph{dry regions} on the solid. The effects of
inertia on the instability have been studied previously in other problems, for
example for a film flowing down an incline~\citep{LMB97}, the breakup of a
liquid filament sitting on a substrate~\citep{ubal_pof2014}, and several other
configurations~\citep{hocking_jfm2002}. However, these problems do not include
explicitly the effects of the intermolecular interaction between the molecules
of the liquid and those of the solid. Here, we consider it by using integrated
Lennard-Jones forces, which lead to the disjoining pressure that entails the
power dependence on the fluid thickness~\citep{kargupta_lang2004}. In the
present context, the occurrence and nature of both inertia and bidimensional
effects in the liquid film on the solid substrate is of interest, not only for
fundamental research, but also for technological applications. 

The solutions of problems under the lubrication approximation is usually limited
to speeds low enough to give small capillary and Reynolds numbers. The extension
of the theory to higher speeds introduces inertia into the problem, and, even in
the case of thin films, the analysis may become much more difficult. The great
simplification previously found by the application of the lubrication theory no
longer exists; instead, the system is governed by the coupling of a nonlinear
partial differential equation for the velocity field, and an evolution equation
for the thickness profile. It is possible, however, to find a class of problems
in which inertial effects can be assessed within the long wave framework. In
this work we are concerned with the instability of a flat liquid film extended
over a solid plane, and subject to intermolecular forces between the liquid and
the solid substrate. Then, the film evolution is studied by considering viscous,
surface tension, and intermolecular forces, with special emphasis on the effect
of inertia in the development of the instability.

\section{Intermolecular forces in the hydrodynamic description}
\begin{figure}
	\centering
	\subfigure[]{\includegraphics[width=.45\textwidth]{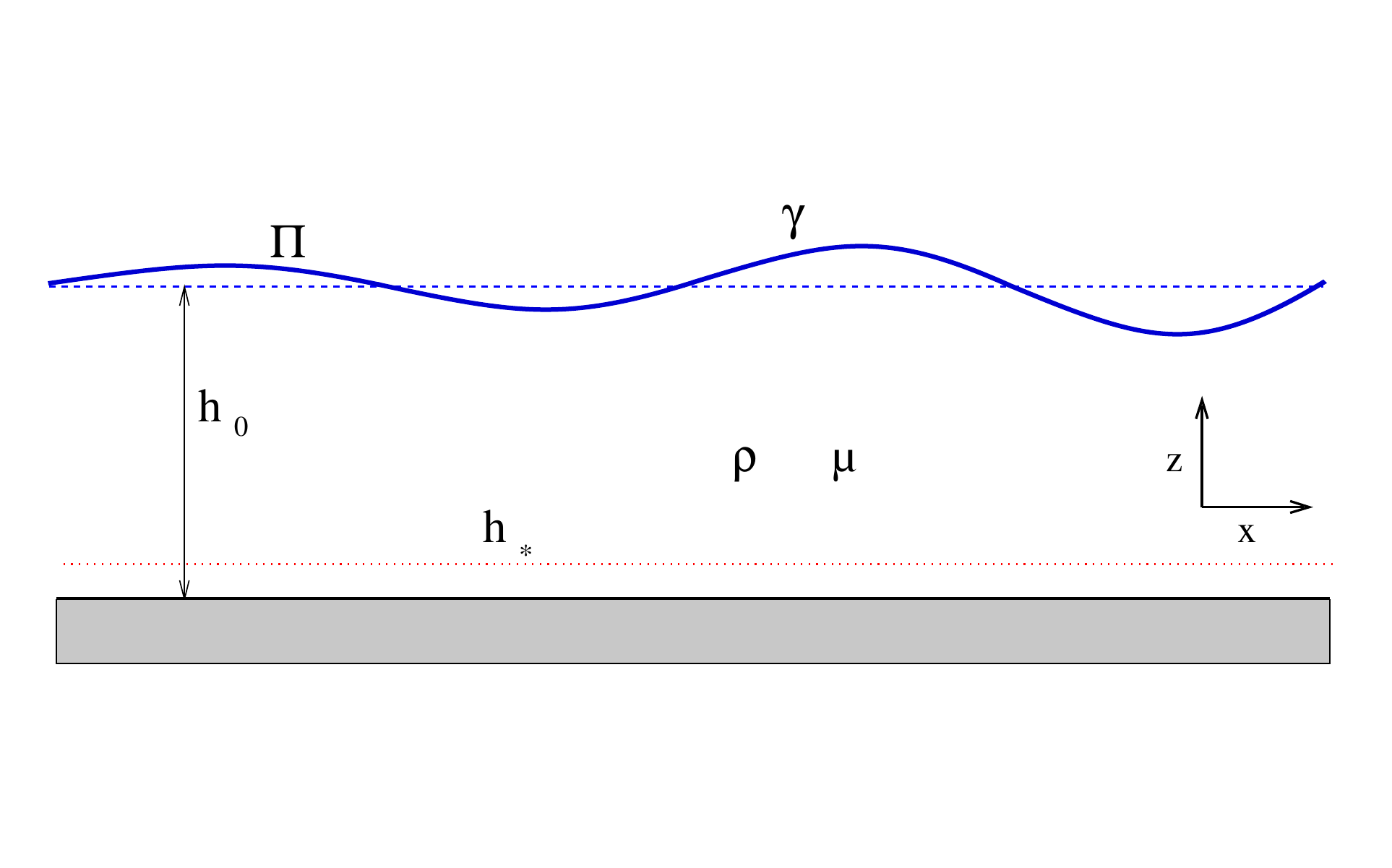}}
	\subfigure[]{\includegraphics[width=.38\textwidth]{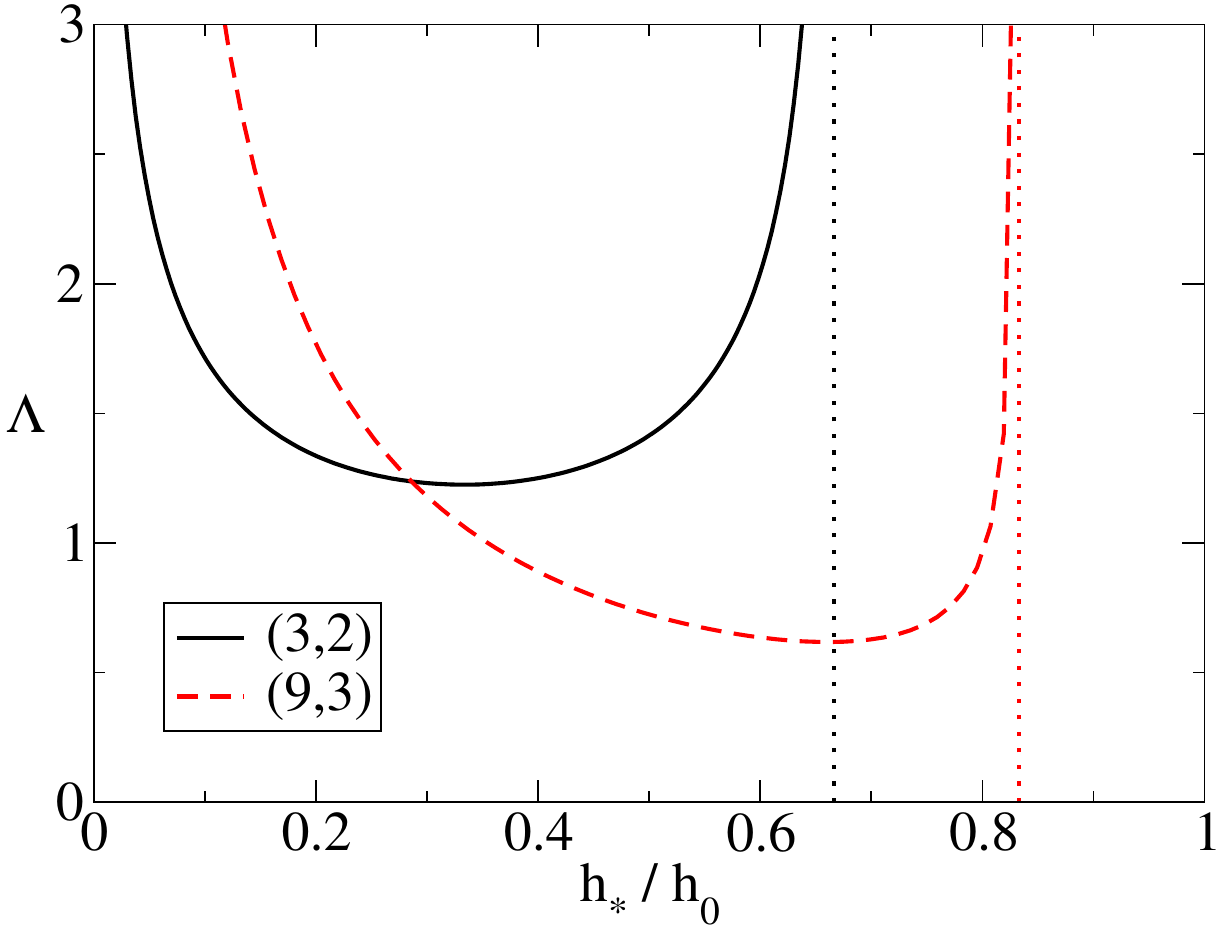}}
	\caption{(a) Schematic diagram of the problem. (b) Parameter $\Lambda$ as given
		by Eq.~(\ref{eq:Lambda}) as a function of the ratio between the equilibrium
		thickness, $h_{\ast}$, and the film thickness, $h_0$, for two pairs of the
		exponents $(n,m)$. The vertical dotted lines correspond to  $g_0=0$, i.e. 
		$h_{\ast}=h_0 \,(m/n)^{1/(n-m)}$.}
	\label{fig:sketch_Lam}
\end{figure}

We consider a thin liquid film of thickness $h_0$, which spans infinitely in the
$x$-direction (the system is invariant in the $y$-direction, i.e. plane flow
conditions prevail), and rests on a solid plane at $z=0$ (see
Fig.~\ref{fig:sketch_Lam}a). Here, we will consider the instability of this
initially flat film under the action of surface tension and intermolecular
forces, acting both at the free surface of the film with instantaneous thickness
$h(x,t)$. Thus, the hydrodynamic evolution is governed by the Navier-Stokes
equation and the incompressibility condition,
\begin{equation}
	\rho \left( \partial _t \mathbf{v} +\mathbf{v} \cdot \nabla \mathbf{v} \right)
	=-\nabla p + \mu \Delta \mathbf{v}, \qquad \nabla \cdot \mathbf{v}=0
	\label{eq:NS_dim}
\end{equation}
where $\rho$ is the liquid density, $\mu$ its viscosity, $p$ the pressure and
$\mathbf{v}= (u,w)$ the velocity field. At the substrate ($z=0$), we apply the
no-slip and non-penetration conditions. At the free surface, $z=h(x,t)$, we have
the usual kinematic condition and normal stress equilibrium given by
\begin{equation}
	p=-\Pi - \gamma \cal{C} 
\end{equation}
where $\gamma$ is the surface tension, $\cal{C}$ the curvature of the surface, and 
\begin{equation}
	\Pi(h)=\kappa  f(h) = \kappa \left[ \left( \frac {h_{\ast}}{h} \right)^n-
	\left( \frac {h_{\ast}}{h} \right)^m \right],
	\label{eq:Pih}
\end{equation}
is the disjoining pressure. Here, $\kappa$ is a constant with units of pressure
(related to the Hamaker constant of the system), the exponents satisfy $n>m>0$,
and $h_{\ast}$ is the equilibrium thickness (of the order of some nanometers).
This surface force is a consequence of the interaction among the molecules in
the three phases present in the problem, namely the liquid of the film, the
solid substrate and the surrounding gas. Note that at equilibrium, i.e. when
$h=h_0=const.$, the film has a uniform pressure $p_0=-\Pi(h_0)>0$, since
$\Pi(h_0)<0$ for $h_0>h_{\ast}$.

\section{Long wave approximation}
In this approximation it is assumed that the film thickness, $h_0$, is much
smaller than the characteristic horizontal length of the problem. Since the
film extends to infinity, we assume that there exists a typical length
associated with the wavelength of the perturbations, namely $\ell$. The
definition of $\ell$ will be made more precise below. Subsequently, for
$\varepsilon = h_0/\ell \ll 1$, we can simplify Eq.~(\ref{eq:NS_dim}) under the
long wave approximation assumptions retaining inertial terms in the form
\begin{eqnarray}
	- \frac{\partial p}{\partial x} + 
	\mu \frac{\partial^2 u}{\partial z^2} &=& \rho \left( \frac{\partial u}{\partial
		t} + u \frac{\partial u}{\partial x} +  w\frac{\partial u}{\partial z} \right)
	\label{eq:mov}\\
	-\frac{\partial p}{\partial z} &=& 0
	\label{eq:dpdz}\\
	\frac{\partial u}{\partial x} + \frac{\partial w}{\partial z} &=&0.
	\label{eq:cont}
\end{eqnarray}

The boundary conditions for these equations are as follows. At $z=0$, we impose
no penetration and no slip at the substrate, 
\begin{equation}
	w=0, \qquad u=0.
	\label{eq:wu0}
\end{equation}
At the liquid-gas interface ($z=h$), we have zero shear stress, 
\begin{equation}
	\frac{\partial u}{\partial z}=0, \qquad 
	\label{eq:shear0}
\end{equation} 
as well as the kinematic condtion,
\begin{equation}
	\frac{\partial h}{\partial t} +u\frac{\partial h}{\partial x} = w,
	\label{eq:kin}
\end{equation} 
and the Laplace relation for the capillary pressure
\begin{equation}
	p(h) =-\gamma \frac{\partial^2 h}{\partial x^2} - \kappa f(h),
	\label{eq:p_h}
\end{equation} 
From Eq.~(\ref{eq:dpdz}) we see that the pressure, $p$, is $z$-independent, and then $p$ 
is only a function of $h$, $p=p(h)$. Thus, we have that the $x$-derivative of $p$ in 
Eq.~(\ref{eq:mov}) is given by
\begin{equation}
	\frac{\partial p}{\partial x}=- \gamma \frac{\partial^3 h}{\partial x^3} -
	\kappa f'(h) \frac{\partial h}{\partial x}.
	\label{eq:dpdx}
\end{equation}

The continuity equation, Eq.~(\ref{eq:cont}), is conveniently satisfied by
introducing the stream function $\psi(x,z,t)$ defined by
\begin{equation}
	u=\frac{\partial \psi}{\partial z}, \qquad w=-\frac{\partial \psi}{\partial x}.
\end{equation}
Therefore, Eqs.~(\ref{eq:mov}) and (\ref{eq:kin}) in terms of $\psi$ are given by
\begin{equation}
	\mu \frac{\partial^3 \psi}{\partial z^3}=- \gamma \frac{\partial^3 h}{\partial
		x^3} - \kappa f'(h) \frac{\partial h}{\partial x} + \rho \left(
	\frac{\partial^2 \psi}{\partial z \partial t} +
	\frac{\partial \psi}{\partial z} \frac{\partial^2 \psi}{\partial x \partial z} -
	\frac{\partial \psi}{\partial x} \frac{\partial^2 \psi}{\partial z^2} \right), 
	\label{eq:psi3_h}
\end{equation}
\begin{equation}
	\frac{\partial h}{\partial t} + \frac{\partial\psi(x,h,t) }{\partial x}=0.
	\label{eq:psi_h}
\end{equation}
The boundary conditions, given by Eqs.~(\ref{eq:wu0}) and (\ref{eq:shear0}), in
terms of $\psi$ are:
\begin{equation}
	\left. \psi \right|_{z=0}=0,\qquad 
	\left. \frac{\partial \psi}{\partial z}\right|_{z=0}=0, \qquad
	\left. \frac{\partial^2 \psi}{\partial z^2}\right|_{z=h} = 0.
\end{equation}

\subsection{Linear stability analysis within long wave approximation}
\label{sec:LSA}
The equilibrium state is given by $h=h_0$, and for small-amplitude perturbations, the
height and stream function can be written in the form
\begin{equation}
	h= h_0 \left( 1+ A e^{\omega t+i k x} \right), \qquad
	\psi= A \psi_1(z) e^{\omega t+i k x},
	\label{eq:lin}
\end{equation}
where $A$ is the small amplitude of the perturbation, and unstable (stable)
modes correspond to $\omega >0$ ($\omega <0$). By replacing Eq.~(\ref{eq:lin})
into Eqs.~(\ref{eq:psi3_h}) and (\ref{eq:psi_h}), and retaining terms up to
order one in $\varepsilon$, we have
\begin{eqnarray}
	&& \mu \frac{d^3 \psi_1}{d z^3}=i \gamma h_0 k^3 - i \kappa h_0 f'(h_0) k +
	\rho \, \omega \frac{d\psi_1}{dz}, \label{eq:psi3}\\
	&& \omega + i k \psi_1(h_0)=0,\label{eq:psi1}
\end{eqnarray}
with the boundary conditions
\begin{equation}
	\left. \psi_1 \right|_{z=0}=0,\qquad \left. \frac{d\psi_1}{dz} \right|_{z=0}=0, 
	\qquad \left. \frac{d^2 \psi_1}{dz^2} \right|_{z=1}=0.
\end{equation}

Now, we define the horizontal length scale, $\ell$, by choosing $\kappa f'(h_0)=
\gamma / \ell^2$, so that it turns out
\begin{equation}
	\ell = \sqrt{\frac{\gamma}{\kappa f'(h_0)}}=
	\sqrt{\frac{\gamma h_0}{\kappa g_0 }},
	\label{eq:ell}
\end{equation}
where 
\begin{equation}
	g_0= h_0 f'(h_0) = - n \left( \frac {h_{\ast}}{h_0} \right)^n +
	m \left( \frac {h_{\ast}}{h_0} \right)^m .
	\label{eq:g0}
\end{equation}
Since $n > m$ and $h_0 > h_{\ast}$, we have $g_0>0$. Note that we are here
including in $\ell$ all the effects related with the magnitude of the
intermolecular forces given by $\kappa$. In fact, this constant is usually
related in the literature with the contact angle, $\theta$, which appears at the
contact regions formed when the film thins up to $h_{\ast}$, and characterizes
the partial wetting of the substrate. It is found that the following simple
relationship holds~\citep{oron_rmp97,schwartz_jcis98,dk_pof07} 
\begin{equation}
	\kappa=\frac {\gamma (1-\cos \theta)}{M h_{\ast}},
\end{equation}
where $M=(n-m)/((n-1)(m-1))$. Thus, the characteristic length scale becomes
\begin{equation}
	\ell = \sqrt{ \frac{M h_0 h_{\ast}}{(1-\cos \theta) g_0}},
\end{equation}
so that this length includes all the parameters determining the problem, except
for $\gamma$ and $\mu$ which yield the time scale (see Eq.~(\ref{eq:tau})
below). In Fig.~\ref{fig:sketch_Lam}b, we show how the dimensionless combination 
\begin{equation}
	\Lambda = \sqrt{1-\cos \theta} \frac{\ell}{h_0} = \sqrt{\frac{M h_{\ast}}{ g_0 h_0}}
	\label{eq:Lambda}
\end{equation}
depends on the ratio $h_{\ast}/h_0$ for two fixed values of the exponents pair
$(n,m)$. Interestingly, very small values of $h_{\ast}$ as well as $h_{\ast}$
close to $h_0 \,(m/n)^{1/(n-m)}$ yield very large values of $\ell/h_0$ for given contact angle,
$\theta < \pi/2$. 

Consequently, a convenient non-dimensional version of the problem for the long
wave approximation is given by the following scaling
\begin{equation}
	X = \frac{x}{\ell},  \quad  Z =\frac{z}{h_0}, \quad H = \frac{h}{h_0}, \quad
	T=\frac{\varepsilon^3}{\tau}t, \quad
	\Psi_1 = \frac{\tau}{\varepsilon^3 \ell} \psi_1, \quad  K = {\ell} k , 
	\quad \Omega  = \frac {\tau}{\varepsilon^3} \omega ,
	\label{eq:scales1D}
\end{equation}
where 
\begin{equation}
	\tau = \frac{\mu \ell}{\gamma}
	\label{eq:tau}
\end{equation}
is the time scale.
Under these definitions, Eqs.~(\ref{eq:psi3}) and (\ref{eq:psi1}) become
\begin{equation}
	\frac{d^3 \Psi_1}{d {Z}^3} + q^2 \frac{d \Psi_1}{dZ}= i {K} ({K}^2 - 1), 
	\label{eq:psi3a}
\end{equation}
\begin{equation}
	\Omega + i {K} \Psi_1(1)=0,\label{eq:psi1a}
\end{equation}
where 
\begin{equation}
	q^2=-La^{\ast} \, \Omega,
	\label{eq:q2}
\end{equation}
with
\begin{equation}
	La^{\ast} = La \, \varepsilon^5 , 
	\label{eq:La_ast}
\end{equation}
and
\begin{equation}
	La = \frac{\rho \gamma \ell}{\mu^2}
	\label{eq:La}
\end{equation}
being the Laplace number. The latter dimensionless number considers the effects of all the
forces playing a role in the flow, namely, inertial (characterized by $\rho$),
viscous (characterized by $\mu$), surface (characterized by $\gamma$) and
intermolecular forces (characterized by $\ell$).

The solution of Eq.~(\ref{eq:psi3a}) has the form
\begin{equation}
	\Psi_1=i K ({K}^2-1) \frac{q Z + \sin (q-q Z) \sec Z - \tan q}{q^3},
\end{equation}
which allows one to obtain the dispersion relation from Eq.~(\ref{eq:psi1a}) as
\begin{equation}
	\frac{\Omega}{{K}^2 \left( {K}^2-1 \right)}=\frac{q-\tan q}{q^3}.
	\label{eq:r1}
\end{equation}
In the limit $q \rightarrow 0$, this expression tends to the purely viscous solution,  
\begin{equation}
	\Omega_{vis}= \frac{{K}^2 (1-{K}^2)}{3},
	\label{eq:w_vis}
\end{equation}
which is obtained in the inertialess case~\citep{dk_pof07} for $La=0$. 
Note that the dimensionless critical (marginal) wavenumber is equal to unity for
the viscous case, i.e. $K_c=1$, so that Eq.~(\ref{eq:w_vis}) shows instability
for $K<1$. This is because the choice of the in-plane characteristic length,
$\ell$, the inverse of the dimensional critical
wavenumber~\citep{nguyen_lang12}.

\begin{figure}
	\centering
	\subfigure[$La^\ast r$ vs. $K$]{\includegraphics[width=.45\textwidth]{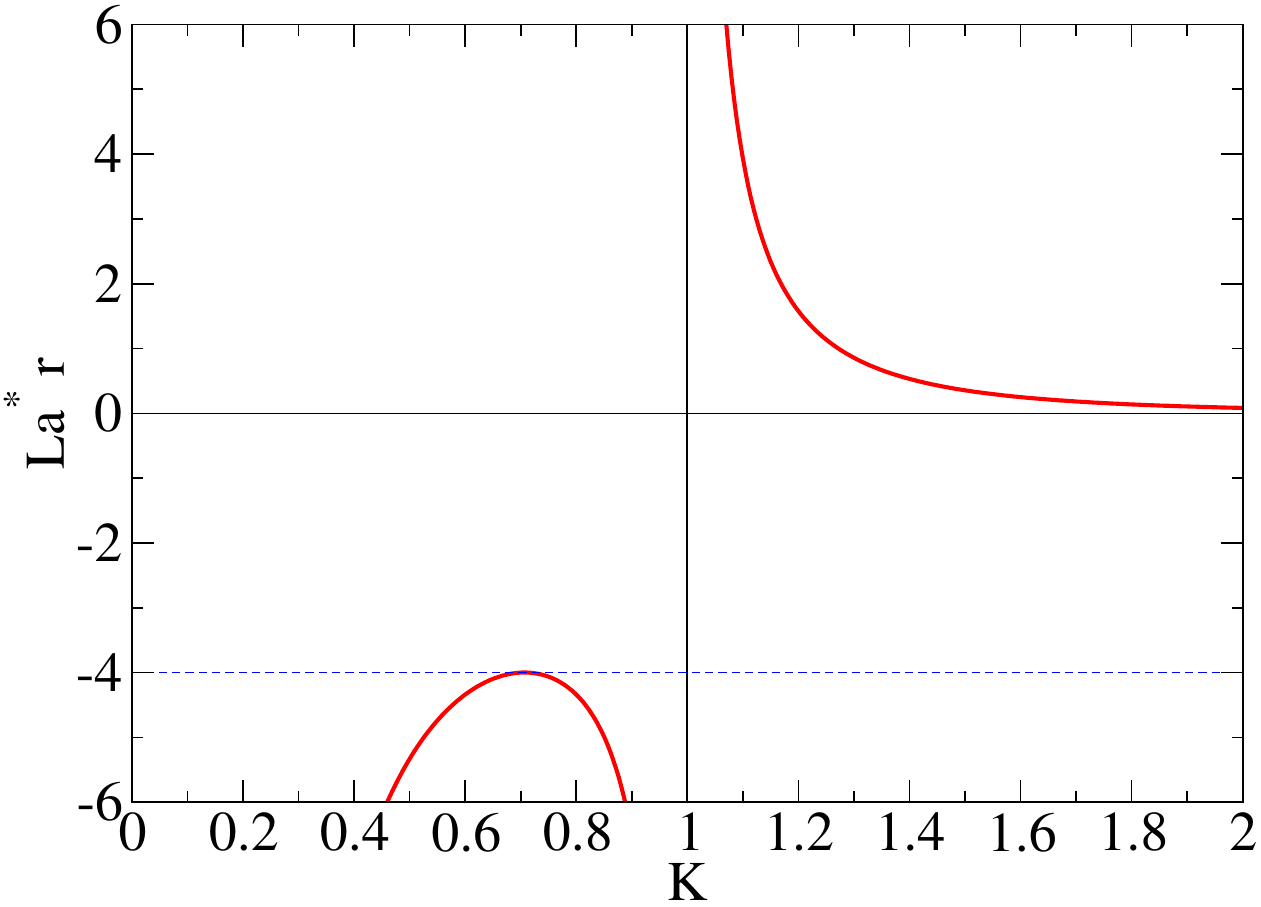}}
	\subfigure[$r$ vs. $q_0$]{\includegraphics[width=.45\textwidth]{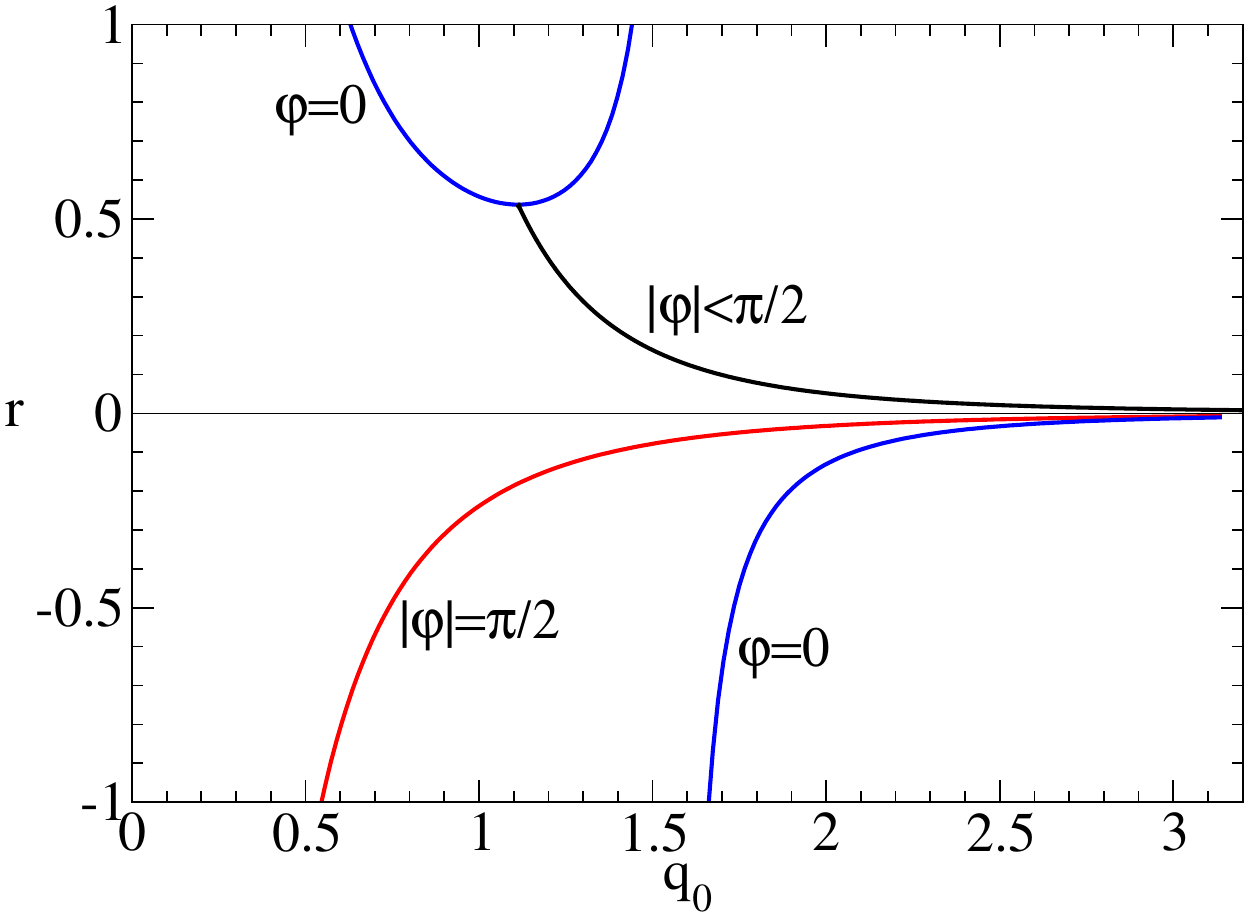}}
	\caption{(a) Relationship between $La^\ast r$ and $K$ as given by
		Eq.~(\ref{eq:rk}). The dashed line is $r=-4/ La^\ast$. (b) Possible values of
		$r$ as a function of $q_0$. The blue lines correspond to $\varphi=0$, the red
		one to $\varphi=\pi/2$ and the black one to $\varphi \neq$ const. (see
		Fig.~\ref{fig:imr0}a).}
	\label{fig:rvskq}
\end{figure}

By dividing Eq.~(\ref{eq:r1}) by $q^2$ and using Eq.~(\ref{eq:q2}), we may
define the parameter $r$ as
\begin{equation}
	r\equiv \frac{1}{La^{\ast} \,{K}^2 ({K}^2-1)},
	\label{eq:rk}
\end{equation}
and then, the possible values of $q$ for given $K$, are given by the roots of 
\begin{equation}
	r=\frac{\tan q - q}{q^5}.
	\label{eq:rq}
\end{equation}
In what follows, we will consider only real values of $K$. Thus, the allowed
values of $r$ are $r<r_{max}=-4/La^\ast$ for $K<1$, and $r>0$ for $K>1$ (see
Eq.~(\ref{eq:rk}) and Fig.~\ref{fig:rvskq}a). In the region $K<1$ and
$r<r_{max}$ there exist two different values of $K$ for a given $r$, and so they
share the same growth rate, $\Omega$. At $r=r_{max}$ we have
$K=K_{m}^{1D}=1/\sqrt{2}$. Instead, in the region $K>1$ and $r>0$, each mode $K$
has a unique and different $r$, and consequently, $\Omega$.

In order to analyse the possible values of $\Omega$ in each region, it is
convenient to introduce the notation
\begin{equation}
	q= q_0 e^{i \varphi},
\end{equation}
so that the complex growth rate is
\begin{equation}
	\Omega = \Omega_r + i \, \Omega_i =-\frac {q_0^2}{La^{\ast}} e^{2 i \varphi}.
	\label{eq:wqphi}
\end{equation}
For $\Omega_r=\Re(\Omega)>0$ ($<0$) we have unstable (stable) modes, and for
$\Omega_i=\Im(\Omega)\neq 0$ we have spatially oscillating modes. Therefore, we
consider the imaginary and real parts of Eq.~(\ref{eq:rq}), which read as
\begin{equation}
	\Re(r)=F(q_0,\varphi)=\left[- \Phi \cos 4 \varphi+
	\sin(2 q_0 \cos\varphi)\cos 5 \varphi +  
	\sinh(2 q_0 \sin\varphi )\sin 5 \varphi  \right]/\Delta, 
	\label{eq:realf}
\end{equation}
\begin{equation}
	\Im(r)=G(q_0,\varphi)= \left[\Phi \sin 4 \varphi-
	\sin(2 q_0 \cos\varphi)\sin 5 \varphi +  
	\sinh(2 q_0 \sin\varphi )\cos 5 \varphi  \right]/\Delta, 
	\label{eq:imf}
\end{equation}
where
\begin{eqnarray}
	\Phi&=&2 q_0 \cos\left(q_0e^{-i \varphi } \right) \cos\left(q_0e^{i \varphi } \right), \\
	\Delta&=&q_0 ^5 (\cos[2 q_0 \cos\varphi]+\cosh[2 q_0 \sin\varphi ]).
\end{eqnarray}

\begin{figure}
	\centering
	\subfigure[]{\includegraphics[width=.45\textwidth]{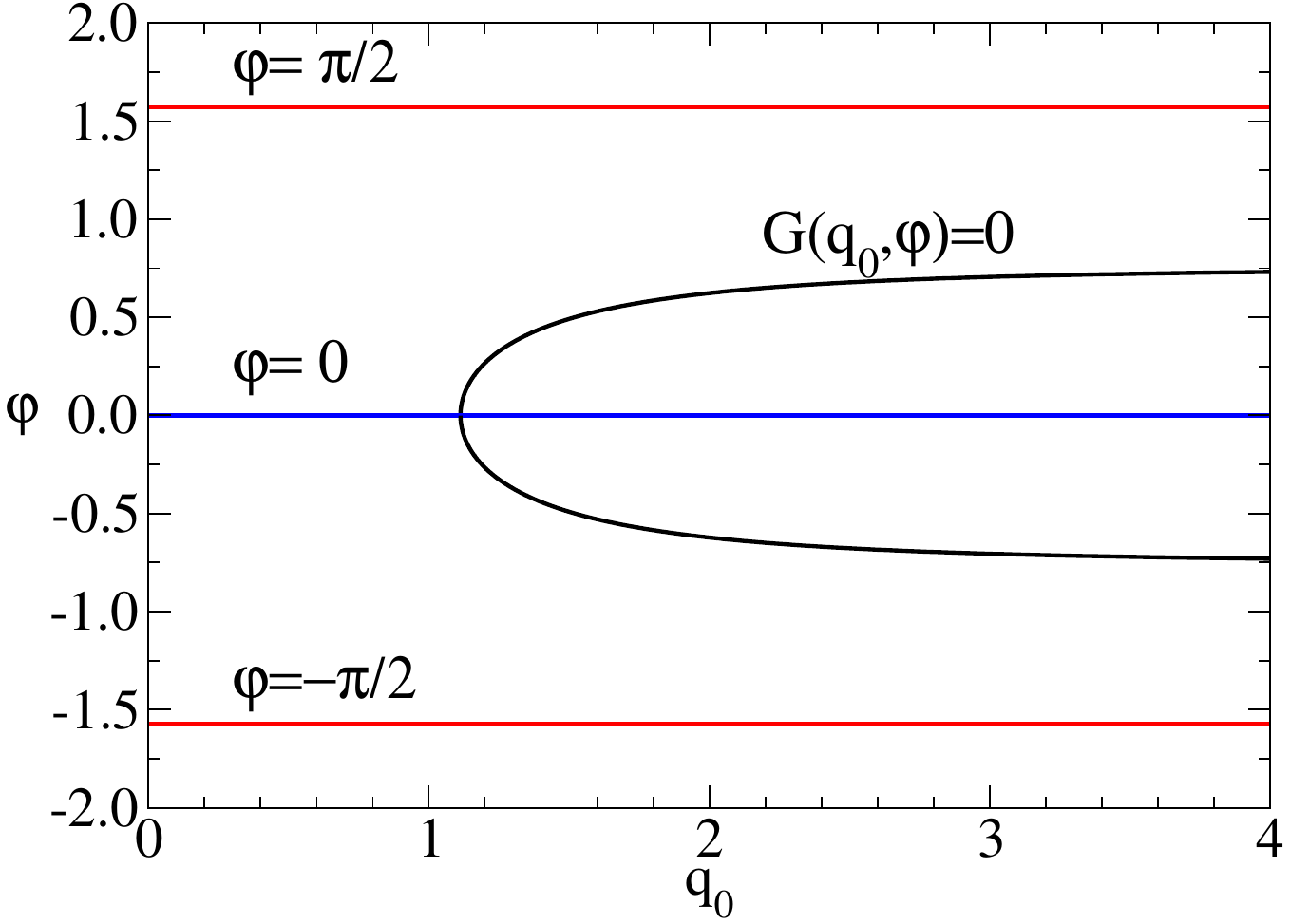}}
	\subfigure[]{\includegraphics[width=.45\textwidth]{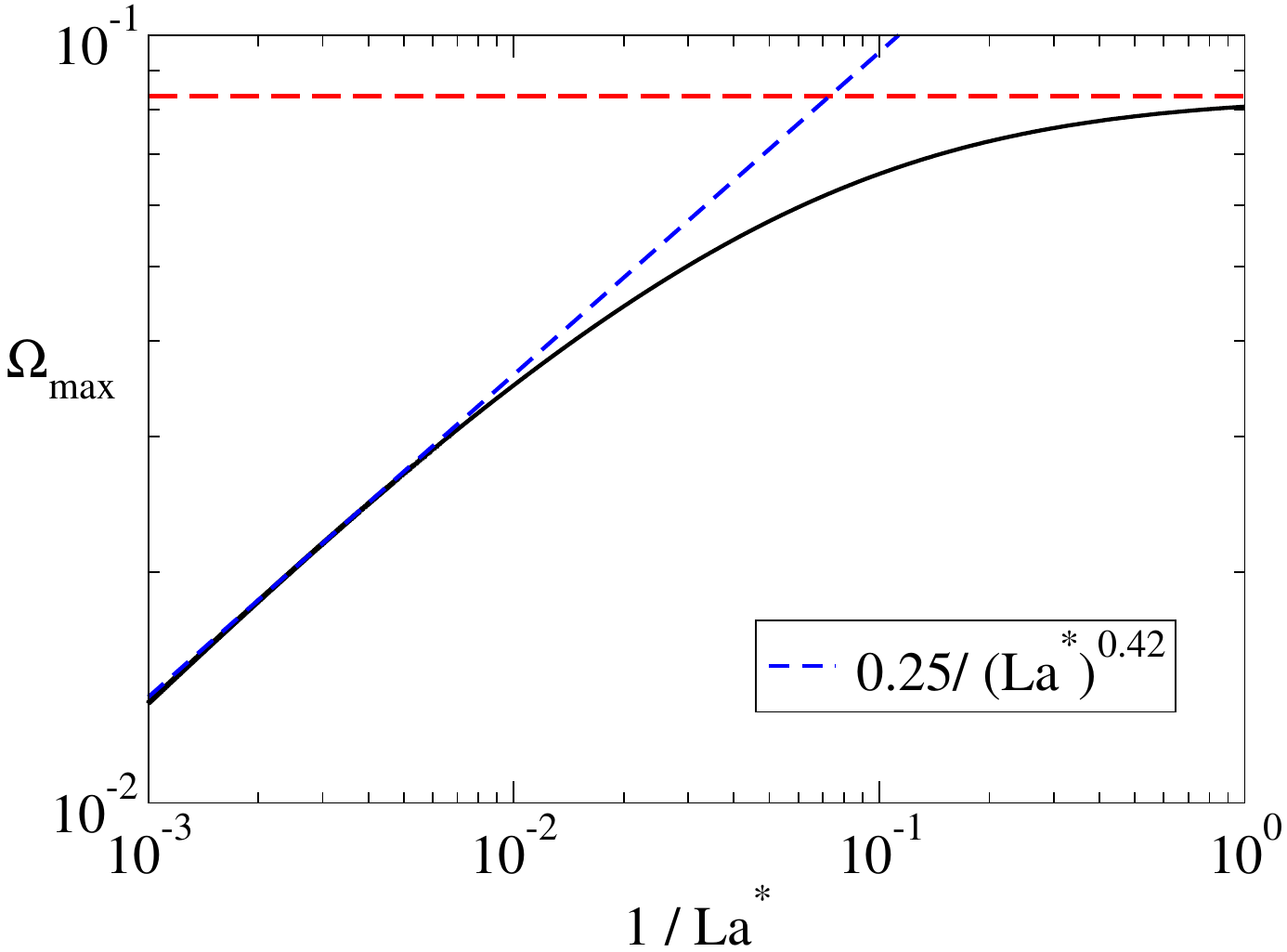}}
	\caption{(a) Curves in the $(q_0,\varphi)$ plane along which $\Im(r)=0$. (b)
		Maximum growth rate as a function of ${La^{\ast}}^{-1}$. The horizontal dashed
		line corresponds to $\Omega_{vis,max}=1/12$.}
	\label{fig:imr0}
\end{figure}

Since $r$ is real, the solutions of Eq.~(\ref{eq:rq}) must have $\Im(r)=0$. Two
trivial roots of this function are $\varphi=0$ and $|\varphi|=\pi/2$. However,
it is possible to find roots also along a curve in the $(q_0,\varphi)$ plane
given implicitly by the function $G(q_0,\varphi)=0$ (see Fig.~\ref{fig:imr0}a).

For $|\varphi|=\pi/2$ we find unstable real modes with growth rates given by
(see Eq.~(\ref{eq:wqphi})) 
\begin{equation} 
	\Omega = \Omega_r =\frac{q_0^2}{La^{\ast}}>0,
\end{equation}
where $q_0$ is now given by the implicit relation
\begin{equation}
	r(\pm i \, q_0)=F(\pm i \, q_0,\pm \pi/2)=\frac{\tanh q_0-q_0}{q_0^5}.
	\label{eq:rq2}
\end{equation}
The function $r(\pm i \, q_0)$ is plotted in with red lines in
Fig.~\ref{fig:rvskq}b. Since $r<0$ for all $q_0$, this branch corresponds to
$K<1$.

Instead, for $\varphi=0$, we obtain stable real modes whose growth rates are
given by (see Eq.~(\ref{eq:wqphi}))
\begin{equation}
	\Omega = \Omega_r =-\frac {q_0^2}{La^{\ast}} <0,
\end{equation}
where $q_0$ is obtained through the implicit relation (see blue lines in
Fig.~\ref{fig:rvskq}b )
\begin{equation}
	r(q_0)=F(q_0,0)=\frac{\tan q_0-q_0}{q_0^5}.
	\label{eq:rq0}
\end{equation}
The function $r(q_0)$ is plotted in with blue lines in Fig.~\ref{fig:rvskq}b. 
Since $r$ changes sign at $q_0=\pi/2$, the upper branch corresponds to $K>1$,
while the lower one to $K<1$. Moreover, these branches are related to
monotonically damped modes. 

The implicit relation $G(q_0,\varphi)=0$ (plotted as the black curve in
Fig~\ref{fig:imr0}a) allows to obtain $r(q_0 e^{i \varphi})$ as a function of
$q_0$ (see black curve in Fig.~\ref{fig:rvskq}b). This branch appears as a
bifurcation point of the the upper branch $\varphi =0$, with coordinates
$B=(r_b,q_{0,b})=(1.1127,0.5367)$. Since $|\varphi| \neq 0, \pm \pi/2$ we have
complex values of the growth rate, $\Omega$, as determined by
Eq.~(\ref{eq:wqphi}). Moreover, since $|\varphi|<\pi/2$, $\Omega_r$ is always
negative and it corresponds to oscillating stable (damped) modes. Besides, it
turns out that $r>0$, so that these modes belong the region $K>1$. 

As a result, only the branch $|\varphi | =\pi/2$ includes unstable modes, which
are in the region $k<1$ and $r<r_{max}$ of Fig.~\ref{fig:rvskq}a. The mode with
maximum (real) growth rate, $\omega_{max}$, is given by $r=r_{max}<0$ for a
given $La^\ast$ and is located at the intersection with the line $|\varphi |=
\pi/2$ in Fig.~\ref{fig:rvskq}b. In fact, for given $La^\ast$ we solve
\begin{equation}
	\frac{\tanh q_{0,max}-q_{0,max}}{q_{max}^5} = -\frac{4}{La^\ast},
\end{equation}
for $q_{0,max}$ and obtain $\Omega_{max}= -q_{0,max}^2 / La^\ast$. The result is
shown in Fig.~\ref{fig:imr0}b, where it is observed how $\Omega_{max}$ tends to
the viscous value, namely $\Omega_{vis,max}=1/12$ (see Eq.(\ref{eq:w_vis})), as
$La^{\ast} \rightarrow 0$. It is also shown that the behavior for large
$La^\ast$ corresponds to a decreasing growth rate as a power law with exponent
close to $0.42$. Similar decreasing trends of the growth rates due to inertial
effects have also been found in other problems~\citep{oron_rmp97,ubal_pof2014}.

Figure~\ref{fig:imr0}b also shows that the line $r=r_{max}$ is also intersected
by the $\varphi=0$ line. Since it corresponds to monotonically damped
perturbations in the region $0<k<1$, this implies that the maximum damping for
the stable mode occurs at the same $k$ than the unstable modes in the $|\varphi
| =\pi/2$.

Note that unstable monotonically growing modes are only possible for $k<1$, so
that neither the critical wavelength nor that of maximum growth rate are
affected by the value of $La^{\ast}$. However, the maximum growth rate itself is
altered by the relative weight of inertial effects with respect to viscosity and
capillarity. Therefore, the Laplace number is relevant when discussing time
scales and growth rates, but not for critical or dominant wavelengths.

The modes with $k>1$ correspond to the $r>0$ region and are always stable as it
is the case in the usual viscous lubrication approximation, but we want now to
analyze whether there is any change in their behaviour when inertia effects are
included. First, note that for each $k>1$, there is a single value of $r>0$ (see
Fig.~\ref{fig:rvskq}a). This value of $r$ could yield either $\varphi=0$ (blue
line, upper branch) or $|\varphi|<\pi/2$ at the black line in
Fig.~\ref{fig:rvskq}b. Two different situations ensue. If $r>r_b$, the solutions
are on the $\varphi=0$ (blue) line, i.e. the modes are monotonically damped, and
two different values of $q$ are admissible: one smaller and the other larger
than $q_{0,b}$. At the point $B=(r_b,q_{0,b})$, both roots degenerate into a
single one. For $0<r<r_b$, the roots are found along the black line, and the
modes are oscillatory and damped. From Eq.~(\ref{eq:rk}), we find the wavenumber
corresponding to point $B$ as given by
\begin{equation}
	K_b=\sqrt{\frac{1}{2}+\sqrt{\frac{1}{4}+\frac{1}{La^{\ast} \, r_b}}}.
	\label{eq:kb}
\end{equation}
Thus, for $1<K<K_b$ there are two damped real modes, while for $K>K_b$ ($r<r_b$)
two oscillatory (complex) damped modes are possible with increasing frequency
oscillations and stronger damping as $K$ increases.

In summary, the condition $\Im(r)=0$ (i.e., $r$ real) yield three types of lines
in the $(q_0,\varphi)$ plane, which can be classified as: 
\begin{enumerate}
	\item $\varphi=0$, which yields stable damped (real) modes, 
	\item $|\varphi|=\pi/2$, which can be related to unstable purely growing (real)
	modes, and \item $\varphi \neq 0, \pi/2$, that will produce stable oscillatory
	modes in time, i.e. complex conjugate roots of $\Omega$.
\end{enumerate}

\begin{figure}
	\centering
	\subfigure[$La^{\ast}=1$]{\includegraphics[width=.45\textwidth]{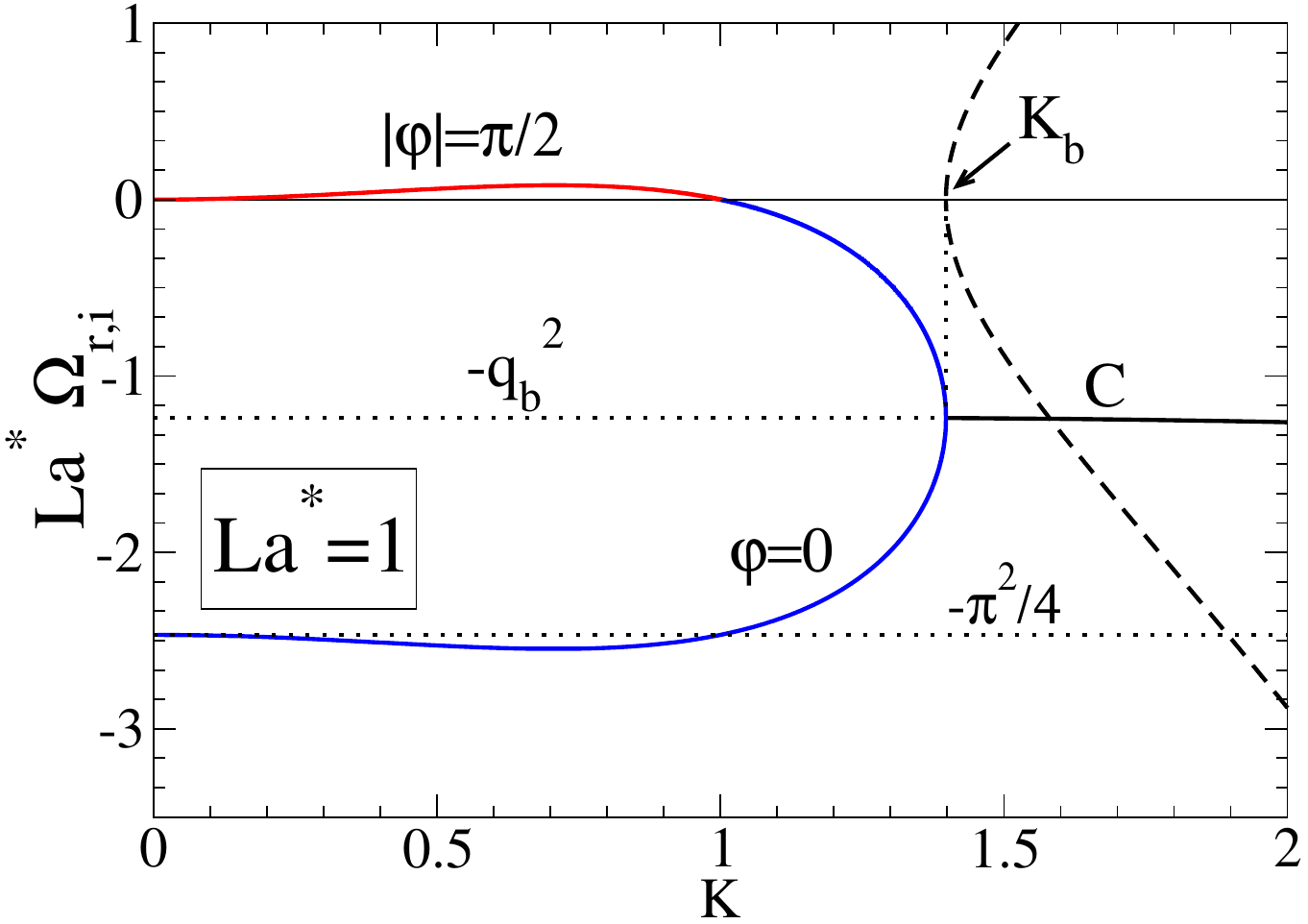}}
	\subfigure[$La^{\ast}=10$]{\includegraphics[width=.45\textwidth]{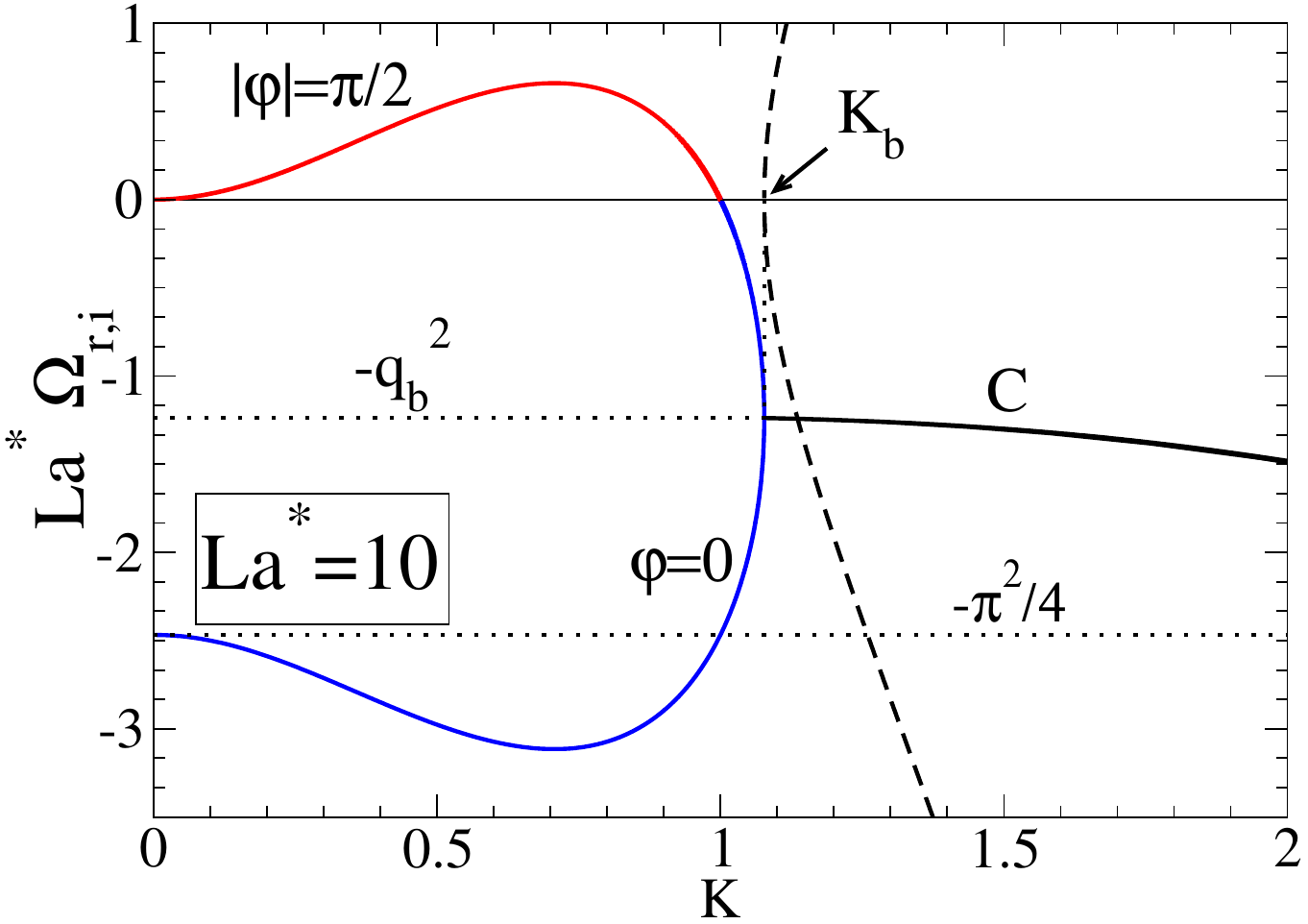}}
	\caption{Real (solid lines) and imaginary (dashed lines) parts of
		$\Omega=\Omega_r+i \Omega_i$ multiplied by $La^\ast$ as a function of the
		wavenumber $K$ for (a) $La^{\ast}=1$, and (b) $La^{\ast}=10$. The curves for
		$\Omega_r >0$ and $K<1$ (unstable region) correspond to $|\varphi|=\pi/2$, and
		those for $\Omega_r < 0$ and $K<K_b$ (stable region for damped modes) 
		correspond to $\varphi =0$}
	\label{fig:omegavsk}
\end{figure}

The procedure to obtain the dispersion relation of the problem, i.e.
$\Omega(K)$, for a fixed $La^\ast$ is as follows. Given a value $K$, we obtain
the corresponding $r$ (see Eq.~(\ref{eq:rk}) and Fig.~\ref{fig:rvskq}a). Then,
with this value of $r$, we find $q_0$ (e.g. using Fig.~\ref{fig:rvskq}b). In the
case of complex roots (black line) the corresponding value of $\varphi$ is a
consequence of requiring that $\Im(r)=0$ in Eq.~(\ref{eq:imf}). Once this is
done, we obtain the full spectrum of modes as shown in Fig.~\ref{fig:omegavsk}.
The dashed lines correspond to $\Omega_i$ for the complex modes along the black
line named C. 

We observe in Fig.~\ref{fig:omegavsk} that $La^{\ast}$ strongly modifies some
features of the complete dispersion relation. For instance, it modifies the
maximum, $\Omega_{max}$, in the unstable region ($K <1$, $\Omega_r>0$). Note
that the product $La^{\ast} \Omega_{max}$ grows with $La^\ast$ because
$\Omega_{max}$ decreases with $La^\ast$ with an exponent less than one (see
Fig.~\ref{fig:imr0}b). Analogously, $La^\ast$ also affects the minimum in the
stable region with $K <1$. For $K >1$, $La^\ast$ only modifies the value of $K_b$
(see Eq.~(\ref{eq:kb})).

\begin{figure}
	\centering
	\subfigure[Unstable modes]{\includegraphics[width=.45\textwidth]{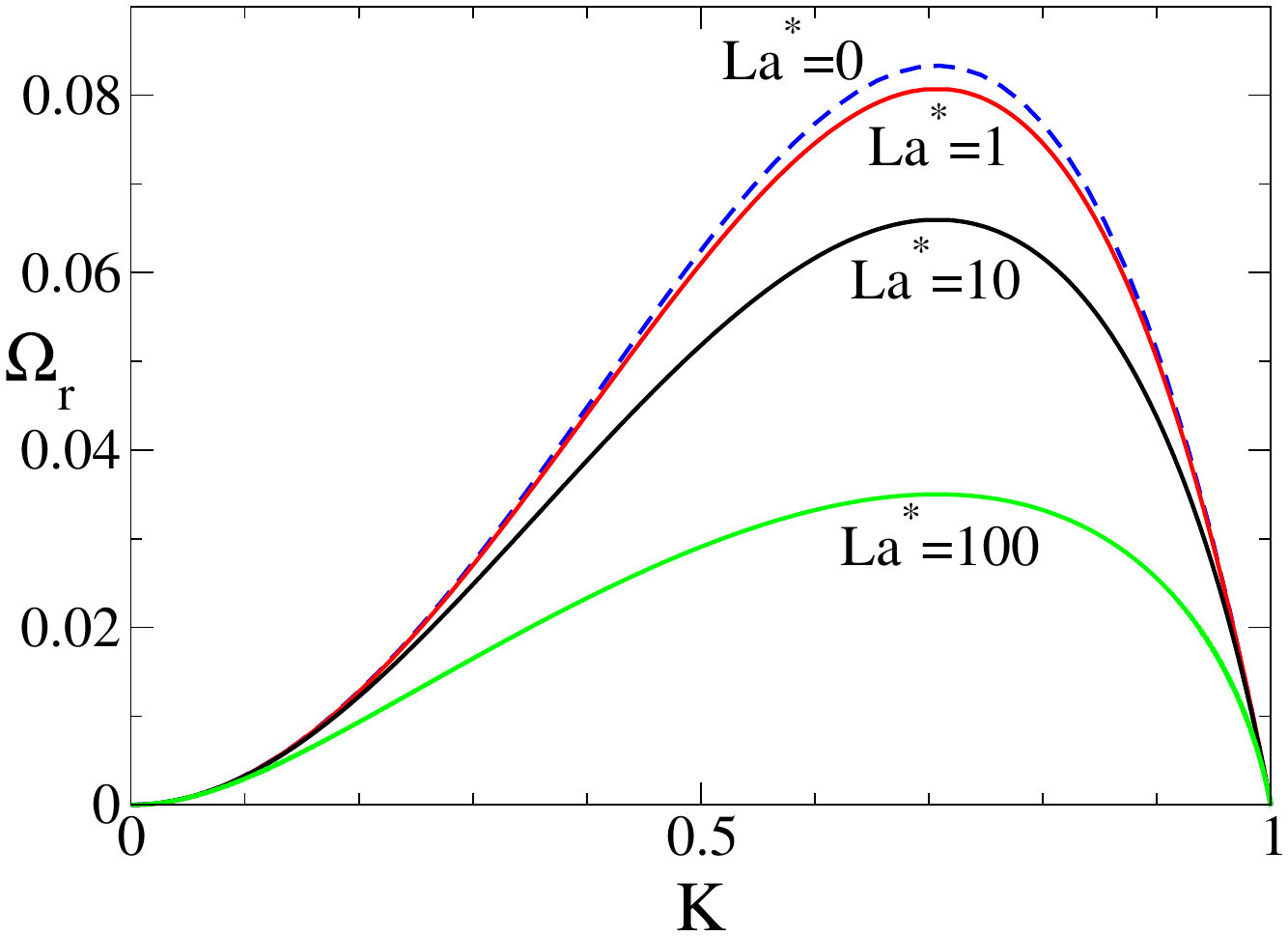}}
	\subfigure[Stable modes]{\includegraphics[width=.45\textwidth]{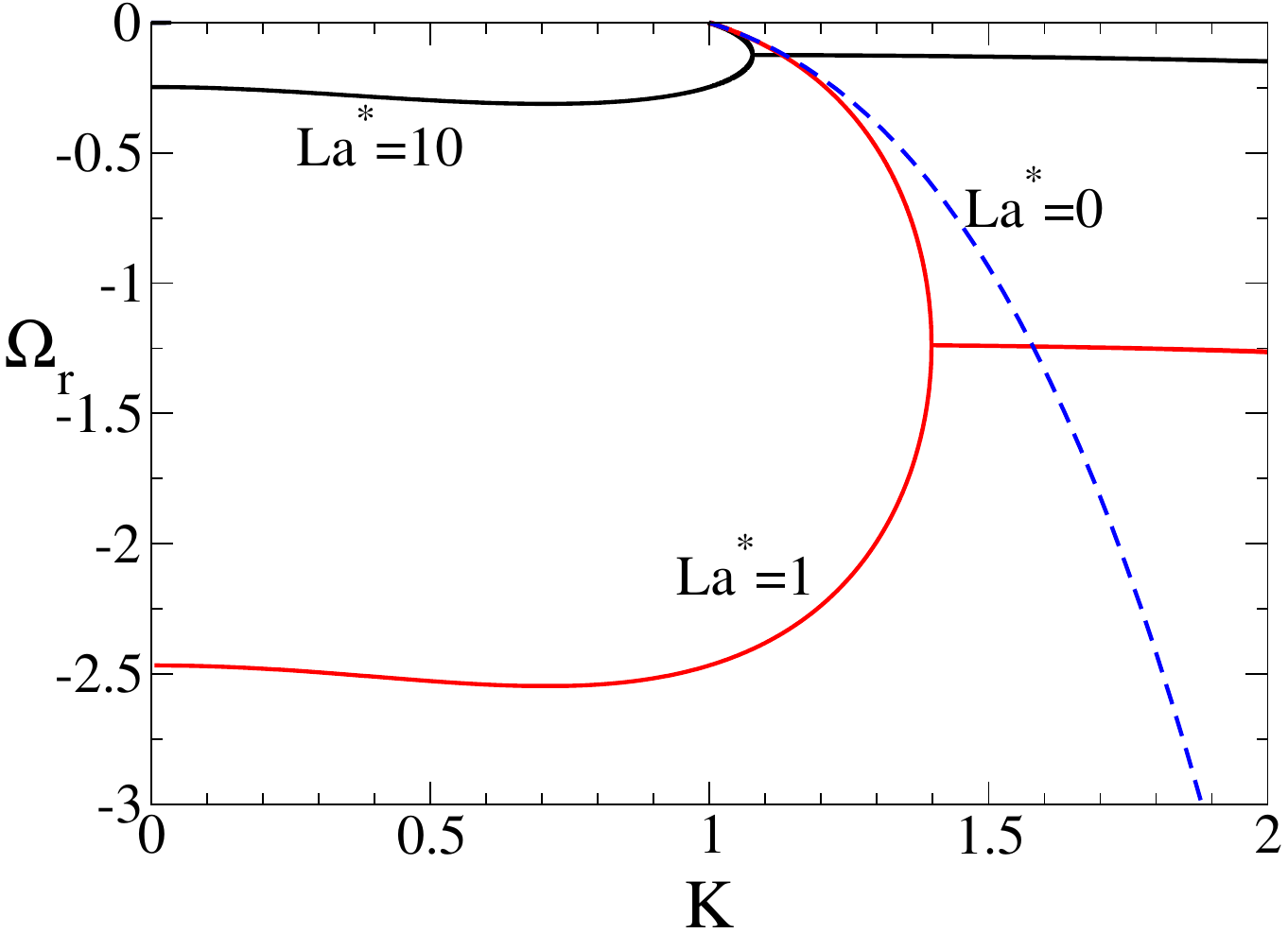}}
	
	\caption{
		Dispersion relations, $\Omega_r(k)$, for some values of $La^{\ast}$: (a) stable
		region, and (b) unstable region. The dashed line for $La^{\ast}=0$ is given by
		Eq.~(\ref{eq:w_vis}).}
	\label{fig:disp}
\end{figure}

In Fig.~\ref{fig:disp} we show a more detailed comparison of the dispersion
curves for several $La^\ast$'s, both on the real growth rates for unstable
($\Omega_r>0$) and stable ($\omega_r<0$) modes. Part a) shows that as
$La^{\ast}$ increases the unstable modes have lower growth rates, but the
wavenumber of the maximum growth is not altered, and remains at
$K_{max}=1/\sqrt{2}$. For very small $La^{\ast}$, the viscous dispersion
relation is rapidly approached (see Eq~(\ref{eq:w_vis})). Figure~\ref{fig:disp}b
shows the stable region of the instability diagram ($K>1$). For $1<K<K_b$),
there two branches of modes that decay exponentially, a characteristic of the
instability which is lost in the viscous approximation. For $K>1$, the viscous
solution, $\Omega_{vis}$, is a fairly good approximation if $K \lesssim K_b$,
but fails for $K$ around $K_b$. Clearly, this solution cannot describe the
oscillating modes for $K >K_b$.

\section{Bidimensional flow: Linear stability analysis}
We consider here the full Navier-Stokes equation, Eq~(\ref{eq:NS_dim}), in its
dimensional form without the long wave approximation assumptions, i.e. the ratio
$\varepsilon$ is not necessarily small now. Therefore, the small perturbations
of the free surface are done on the velocity and pressure fields, and are
expressed in terms of normal modes with a wavenumber $\mathbf{k} = (k_x,0,0)$
parallel to the substrate. Thus, we have
\begin{eqnarray}
	\delta \mathbf{v}&=& \mathbf{v}(z) \, e^{i \mathbf{k}\cdot\mathbf{r}+\omega t},\nonumber \\
	p&=&p_0 + \delta p = p_0 + p_1(z) \,e^{i \mathbf{ k}\cdot\mathbf{r}+\omega t}, \\
	h&=& h_0 + \delta h = h_0 + \zeta \, e^{i \mathbf{k}\cdot\mathbf{r}+\omega t}, \nonumber
	\label{eq:pert}
\end{eqnarray}
where $\mathbf{v}=(u(z),0,w(z))$ and $\delta h$ is the Lagrangian displacement
of the free surface. Note that, for small perturbations, we have $\zeta=
w(1)/\omega$. Then, the Navier--Stokes equation at first order in the
perturbations becomes
\begin{equation}
	\rho \, \partial _t \delta \mathbf{v}=-\nabla \delta p + \mu \Delta \delta \mathbf{v}.
	\label{eq:NS_1}
\end{equation}

Since we assume incompressible flows, $i \mathbf k \cdot \mathbf v=-D w$, where
$D\equiv d/dz$. In order to reduce the number of variables, we eliminate the
pressure terms, by taking the $z$ component of the
$\nabla\times\nabla\times$Eq.~(\ref{eq:NS_1}). After some calculations, we
obtain
\begin{equation}
	(D^2-k^2_x)(D^2-s^2\, k^2)w=0
	\label{eq:2Dlin}
\end{equation}
where $k=k_x$, and 
\begin{equation}
	s^2 \equiv 1+  \, \omega / ( \nu k^2), 
	\label{eq:s2}
\end{equation}
or equivalently
\begin{equation}
	\omega =\left(s^2-1\right)\nu k^2 .
	\label{eq:s_w}
\end{equation}
The general solution of Eq~(\ref{eq:2Dlin}) is
\begin{equation}
	w = A_1 \cosh (k z) + A_2\cosh (s\, k z) +  A_3 \sinh (k z)+ A_4 \sinh(s\, k z),
\end{equation}
where the constants $A_i$ ($i=1,...,4$) are calculated by applying the following
boundary conditions.

First, we impose the no flow condition through the rigid substrate, 
\begin{equation}
	\left. w \right|_{z=0}=0.
	\label{eq:w0}
\end{equation}

Second, we shall assume that there is no slip at the substrate, 
$\left. u\right|_{z=0}=0$. Since the flow is incompressible, then 
\begin{equation}
	\left. D w\right|_{z=0}=0
	\label{eq:Dw0}
\end{equation}

Third, the tangential stresses at the free surface should be zero,
$\left.\mathbf{k}\cdot {\cal S} \cdot \mathbf{e}_z\right|_{z=h_0}=0$, where
${\cal S}=-p \mathbf I + \mu \nabla \delta \mathbf{v} + \mu (\nabla \delta
\mathbf{v} )^T $ is the stress tensor. By replacing here the perturbed
quantities, Eq.~(\ref{eq:pert}), we find
\begin{equation}
	\left.( D^2+k^2)w\right|_{z=h_0}=0,
	\label{eq:esft}
\end{equation}

Finally, the normal stress at the free surface must satisfy the generalized
Laplace pressure jump,
\begin{equation}
	\left. \mathbf{e}_z\cdot {\cal S} \cdot \mathbf{e}_z \right|_{z=h_0}=\gamma {\cal C}+\Pi,
\end{equation}
where ${\cal C}=- k^2 \zeta$ is the first order curvature of the perturbed free
surface. Since 
\begin{equation}
	\zeta=\left. \frac {w}{\omega}\right|_{z=h_0},
\end{equation} 
we have
\begin{equation}
	\left. ( -p_1  + 2 \mu D w ) \right|_{z=h_0}=
	\left. \frac{d \Pi}{dh}\right|_{h=h_0}\zeta -\gamma k^2 \zeta
	\label{eq:esfn}
\end{equation}
Notice that the term in $d \Pi/dh$ plays a role that is analogous to that of
$\rho g$ in the Rayleigh--Taylor instability of a thin film. In order to obtain
$p_1$ for this equation, we perform the scalar product of Eq.~(\ref{eq:NS_1}) by
$\mathbf k$, and using Eq.~(\ref{eq:s_w}) we find
\begin{equation}
	p_1 =\mu \left(\frac{D^2}{k^2}-s^2 \right)D w.
\end{equation}
Then, by replacing this expression of $p_1$ at $z=h_0$ into Eq.~(\ref{eq:esfn}),
we finally write this boundary condition as
\begin{equation}
	\left. \frac{\gamma}{\ell^2} (k^2 \ell^2- 1) \frac {w}{\omega}\right|_{z=h_0}
	=\left. \mu \left(\frac{D^2}{k^2}-2-s^2 \right) Dw \right|_{z=h_0}
	\label{eq:esfn1}
\end{equation}
where $\ell$ is defined by Eq.~(\ref{eq:ell}).

From the above boundary conditions, Eqs.~(\ref{eq:w0}), (\ref{eq:Dw0}),
(\ref{eq:esft}) and (\ref{eq:esfn1}), it is possible to build up a matricial
system to solve the four unknowns, $A_i$ ($i=1,...,4$). Its determinant must be
zero to avoid a trivial solution. This condition leads to 
\begin{eqnarray}
	&{K}^3 \left[-4 \left( s+s^3 \right)
	+s \left( 5+2 s^2+s^4 \right) \cosh({K} \varepsilon) \cosh(s\,{K} \varepsilon)- 
	\right. \nonumber \\
	& \left. \left( 1+6 s^2+s^4\right) \sinh({K} \varepsilon) 
	\sinh(s\, {K} \varepsilon) \right]+  \nonumber \\
	&La \left({K}^2-1\right) \left[s \cosh(s\, {K} \varepsilon) \sinh({K} \varepsilon)  
	-  \cosh({K} \varepsilon ) \sinh(s\, {K} \varepsilon ) \right] =0,
	\label{eq:DR} 
\end{eqnarray}
with $K$ and $La$ defined in Eqs.~(\ref{eq:scales1D}) and (\ref{eq:La}),
respectively. 

This expression is the dispersion relation of the problem, since it implicitly
gives $s$ as a function of $K$. The values of $\omega$ can be obtained through
Eq.~(\ref{eq:s_w}), which in dimensionless variables is
\begin{equation}
	\Omega \, = \omega \tau =\frac{{K}^2}{\varepsilon^3 La}(s^2-1).
	\label{eq:Omega}
\end{equation}
It can be shown that Eq.~(\ref{eq:DR}) is identical to that obtained in
\cite{kargupta_lang2004} if slipping at the substrate is neglected once we take
into account that $\alpha$ and $\beta$ in their Eqs.~(6) and (7) are 
$K \varepsilon$ and $s\, K \varepsilon $ respectively (our $s$ corresponds to 
their $q$). 

In order to obtain the limit of Eq.~(\ref{eq:DR}) for $\varepsilon \ll 1$, note
first that $q$ in Eq.~(\ref{eq:q2}) of the long wave model is related to $s$ in
Eq.~(\ref{eq:s2}) by
\begin{equation}
	q^2=(1-s^2) {K}^2 \varepsilon^2,
	\label{eq:qsbeta}
\end{equation}
and that $K \, \varepsilon = 2 \pi h_0/\lambda << 1$ in this limit. In order to
keep a meaningful value of $q$, $|s| \gg 1$ is required, which means that
$q\approx i K s \varepsilon$. Thus, with these ingredients in mind when
analysing the limiting behavior of the dispersion relation given by
Eq.~(\ref{eq:DR}), to the lowest meaningful order in $\varepsilon$, we find 
\begin{equation}
	La^{\ast}\, {K}^2({K}^2-1)(\tan q - q)=q^5
	\label{eq:DRlim}
\end{equation}
which is the same expression as given by the long wave model when
Eqs.~(\ref{eq:rk}) and (\ref{eq:rq}) are combined.

\section{Comparison between long wave and bidimensional models}
In this section we study the effects of $La$ and $\varepsilon$ on the dispersion
relation for the unstable region as given by the one dimensional (1D) long wave
approximation and the bidimensional (2D) model. For the 1D case, we focus on the
solution of Eqs.~(\ref{eq:rk}) and (\ref{eq:rq}) for $\varphi=\pi/2$, while for
the 2D case we numerically solve Eq.~(\ref{eq:DR}) together with
Eq.~(\ref{eq:Omega}).

\begin{figure*}
	\centering
	\includegraphics[width=.95\textwidth]{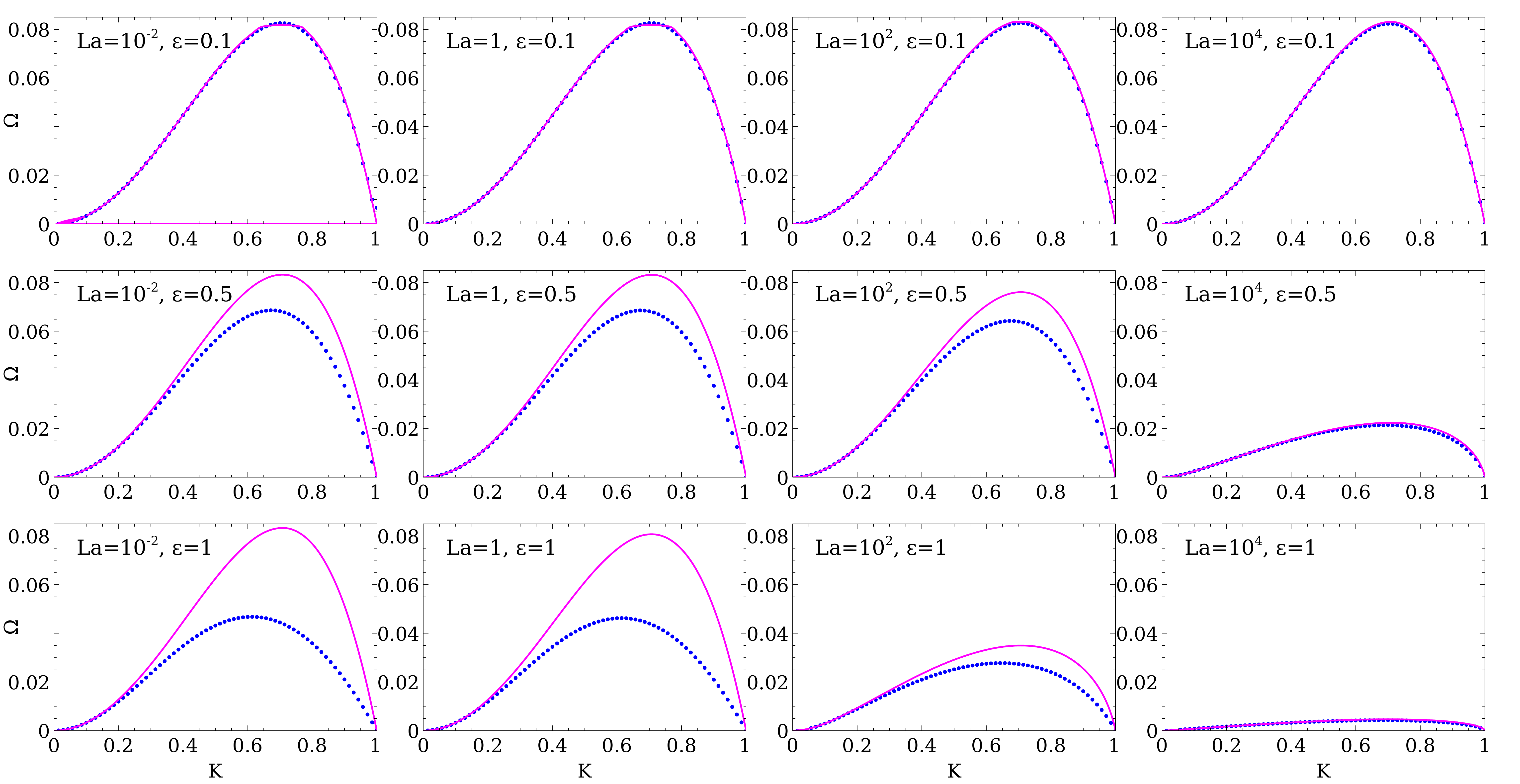}
	\caption{Dispersion relations, $\Omega$ as a function of $K$, for different
		values of $La$ and $\varepsilon$ for the linearized problem: 2D model (N--S
		solution, blue dots) and 1D long wave approximation with inertia (purple lines).
		For large $La$ (strong inertia) we obtain similar results for both models. For
		small $La$ (weak inertial effects) there are meaningful differences between both
		models (different $\varepsilon$'s)}
	\label{fig:RD_seq}
\end{figure*}

In Fig.~\ref{fig:RD_seq} we show the comparison between 1D and 2D dispersion
relations for given values of $La$ (columns) and $\varepsilon$ (rows). The
inertial effects are shown along a given row (fixed $\varepsilon$), with the
first column being a viscous dominated flow, and the fourth column corresponding
to inertia dominated cases. For small $\varepsilon$, as in first row where
$\varepsilon=0.1$, both dispersion relations are practically coincident for any
value of $La$, as expected and shown analytically in Eq.~(\ref{eq:DRlim}). In
general, the long wave model qualitatively predicts the same trends as the 2D
model. However, for $\varepsilon$ as large as $\varepsilon=0.5$ (second row),
the quantitative comparison certainly depends on $La$: the smaller $La$, the
larger is the departure between both models, i.e. 2D effects become more
important for flows with weak inertia. This effect is still more pronounced for
larger $\varepsilon$ as seen in the third row for $\varepsilon=1$. Also note
that, for fixed $La$, the position of the maximum shifts towards the left as
$\varepsilon$ increases. 

We focus now on the behavior of the maximum of the dispersion relations, since
its analysis provides interesting insight on the effects of both inertia and
aspect ratio. While the behavior for the 1D model has been already described, 
the 2D model results can be obtained noticing that for $\Omega=\Omega_{max}$:
\begin{equation}
	0=\frac{d\Omega(s,K)}{dK}=\frac{\partial \Omega}{\partial K} + \frac{\partial \Omega}{\partial s} \frac{ds}{dK}.
\end{equation}
Since the dispersion relation satisfies $F(s,K)=0$, one can calculate
\begin{equation}
	\frac{ds}{dK}=-\frac{\frac{\partial F}{\partial K}}{\frac{\partial F}{\partial s}}.
\end{equation}
Thus, using Eq.~(\ref{eq:Omega}), it is possible to write
\begin{equation}
	2 K (s^2-1)\frac{\partial F}{\partial s} + 2 s K^2 \frac{\partial F}{\partial K}=0,
\end{equation}
which we shall not write in full for brevity. By solving this expression in conjunction with  
Eq.~(\ref{eq:DR}) we are able to obtain $\Omega_m$ and $K_m$ as a function of both 
$La$ and $\varepsilon$. 

\begin{figure}
	\centering
	\subfigure[] {\includegraphics[width=.45\textwidth]{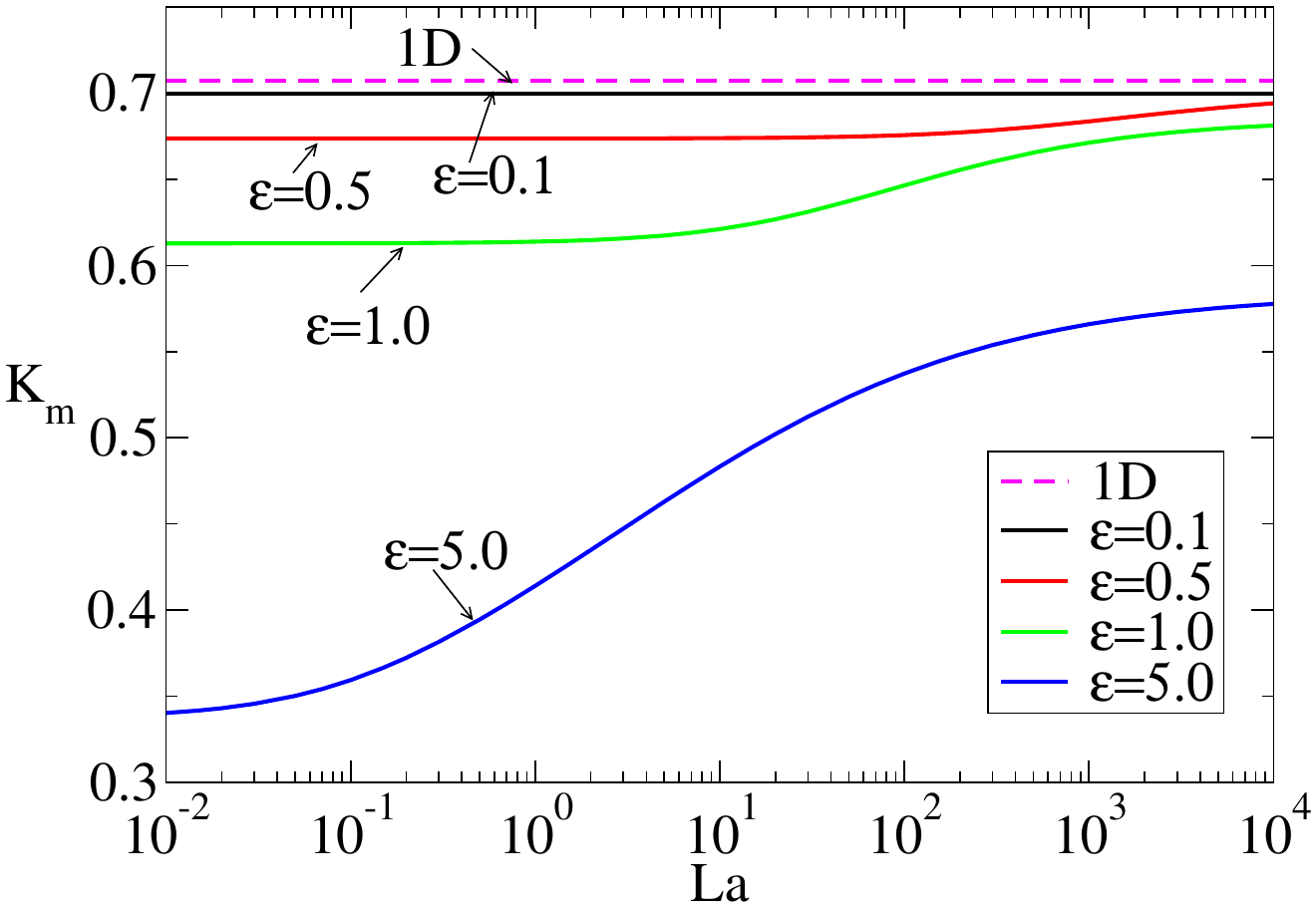}}
	\subfigure[] {\includegraphics[width=.45\textwidth]{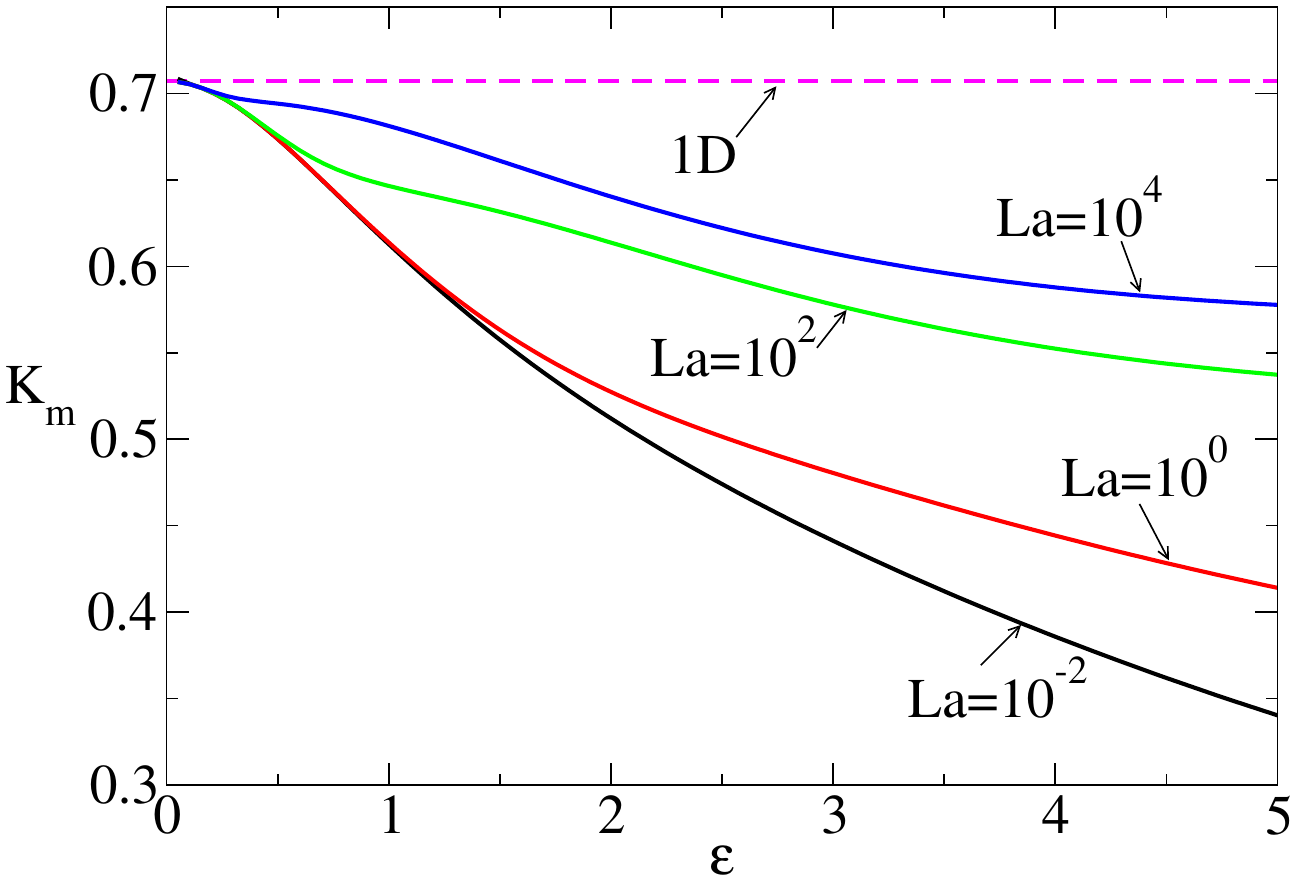}}
	\caption{
		Wavenumber at the maximum growth rate, $K_m$, as a function of: (a) $La$ for
		different $\varepsilon$'s, and (b) $\varepsilon$ for different $La$'s. The solid
		lines corresponds to 2D model, and the dashed line to the 1D (long wave) model,
		$K_m^{1D}=1/\sqrt{2}=0.707$.}
	\label{fig:km}
\end{figure}

In Fig.~\ref{fig:km} we show the wavenumber at the maximum growth
rate, $K_m$, as a function of $La$ for several aspect ratios $\varepsilon$'s,
and vice-versa. Recall that for 1D model, we simply have $K_{m}=1/\sqrt{2}$,
independently of both $La$ and $\varepsilon$. For small $La$, the departure
between both models can be very large if $\varepsilon$ is not very small. In
fact, the value of $K_m$ can be reduced even up to $50\%$ for $\varepsilon$ as
large as $\varepsilon =5$ (see Fig.~\ref{fig:km}a). The difference remains also
for large $La$, but it reduces for smaller $\varepsilon$. This effect is clearly
shown in Fig.~\ref{fig:km}b since, even if the departure increases for
$\varepsilon$ increasing, it is smaller for larger $La$'s. Therefore, the
lubrication and the long wave approximations predict pretty larger distances
between drops after breakup if the corresponding aspect ratio does not fulfill
the requirement $\varepsilon \ll 1$. However, this discrepancy is smaller for
larger $La$'s.

\begin{figure*}
	\centering
	\subfigure[] {\includegraphics[width=.45\textwidth]{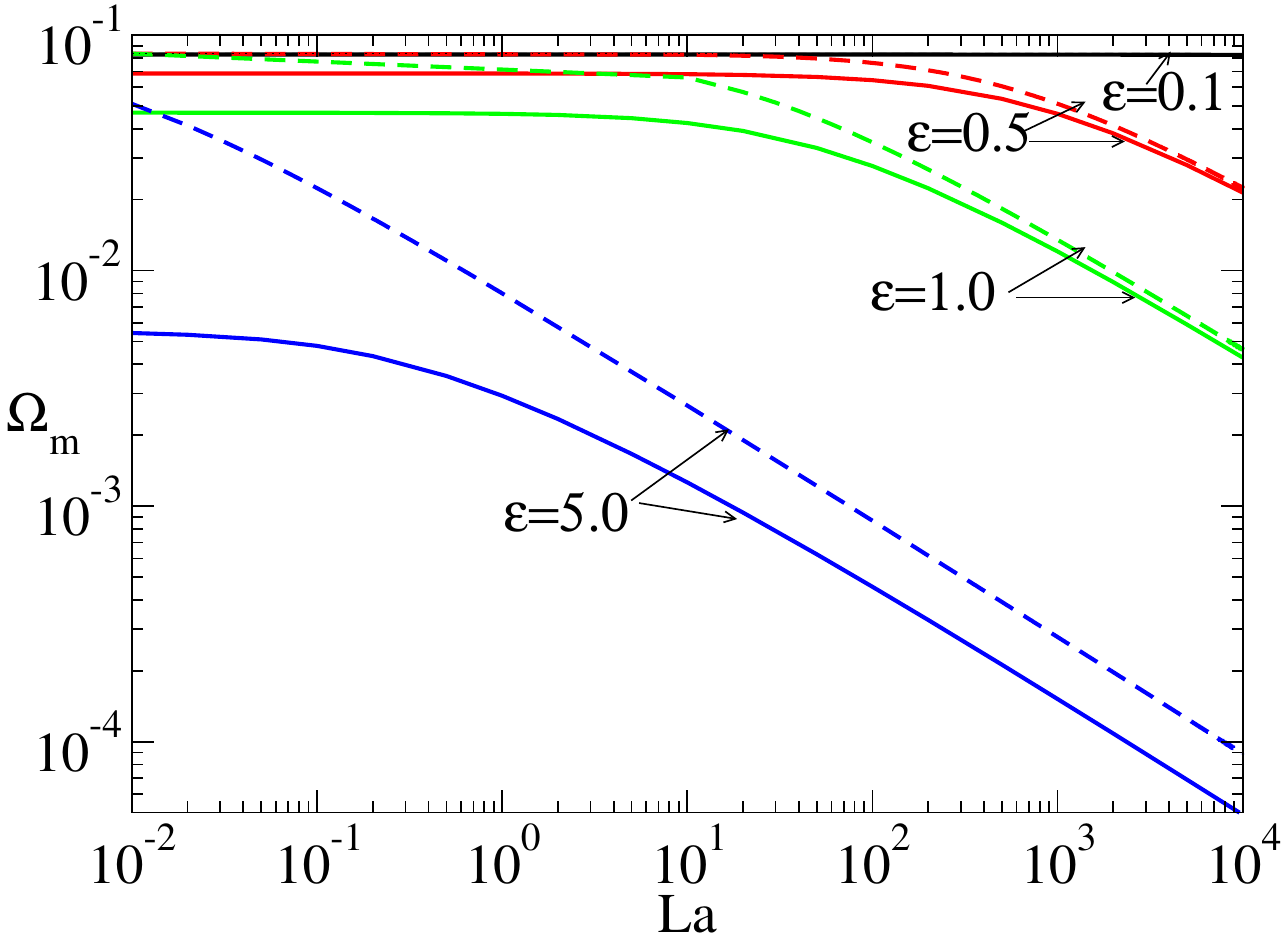}}
	\subfigure[] {\includegraphics[width=.45\textwidth]{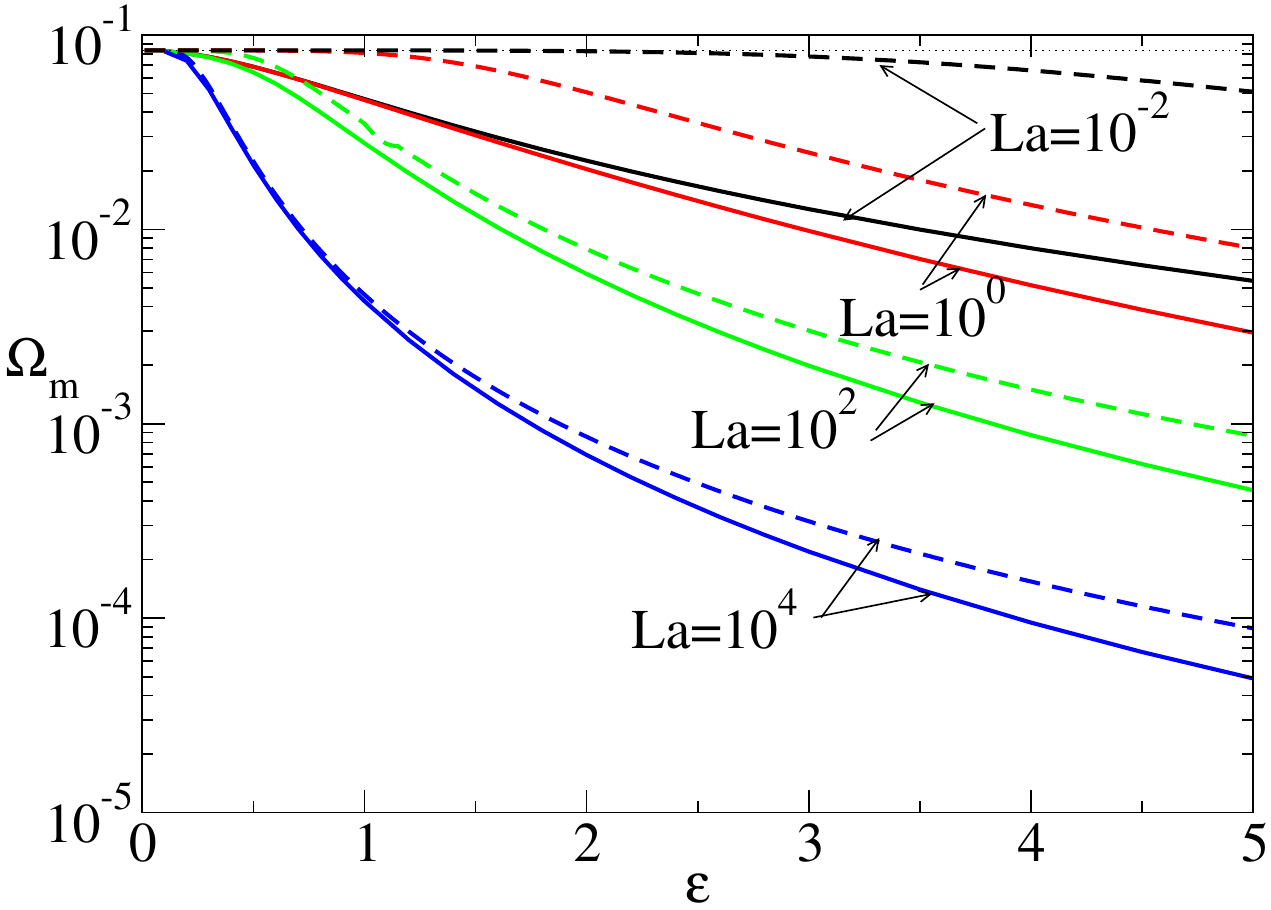}}
	\caption{
		Maximum growth rates, $\Omega_m$, as a function of: (a) $La$ for different
		$\varepsilon$'s, and (b) $\varepsilon$ for different $La$'s. The solid lines
		corresponds to 2D model, and the dashed line to the 1D (long wave) model (same
		colour implies same $\varepsilon$ in (a), and same $La$ in (b)). The upper
		dotted line in both figures is the purely viscous ($La=0$) growth rate,
		$\Omega_{m,vis}=1/12=0.0833$. In (a), the 1D and 2D models for $\varepsilon
		=0.1$ are graphically superimposed to the this value of $\Omega_m$.}
	\label{fig:wm}
\end{figure*}

Figure~\ref{fig:wm} shows the maximum growth rate, $\Omega_m$, as a function of
$La$ for several aspect ratios $\varepsilon$'s, and vice-versa. The curves for
1D model are obtained for the corresponding value of $La^\ast$ as given by
Eq.~(\ref{eq:La_ast}). The difference in $\Omega_m$ between both models for
small $La$ can be very large if $\varepsilon$ is sufficiently large (see
Fig.~\ref{fig:wm}a). Instead, for large $La$ both models agree in a power law
decrease of $\Omega_m$ with differences that increase for larger $\varepsilon$,
as expected. The departures for small $La$ and large $\varepsilon$ are also seen
in Fig.~\ref{fig:wm}b which shows how the 1D curves with smaller $La$ separate
more and more from the corresponding 2D ones for smaller $La$'s. Thus, for small
values of $La$ the discrepancy in $\Omega_m$ between 1D and 2D models can be
very large even if $\varepsilon$ is not strictly much less than one (see also
Fig.~\ref{fig:ampl_seq}).

\section{Numerical simulations}
In order to analyse the validity range of the predictions of both LSA's
described above, we perform numerical simulations of the instability by solving
the complete set of Navier-Stokes equations. Here, we use the two-phase flow,
moving mesh interface of COMSOL Multiphysics. It solves the full
incompressible Navier--Stokes equations using the Finite Element technique in a
domain which deforms with the moving fluid interface by using the Arbitrary
Lagrangian-Eulerian (ALE) formulation. The interface displacement is smoothly
propagated throughout the domain mesh using the Winslow smoothing algorithm. The
main advantage of this technique compared to others such as the Level Set of
Phase Field techniques is that the fluid interface is and remains sharp. The
main drawback, on the other hand, is that the mesh connectivity must remain the
same, which precludes the modelling of situations for which the topology might
change. The default mesh used throughout is unstructured and has $2940$
triangular elements (P1 linear elements for both velocity and pressure).
Automatic remeshing is enabled to allow the solution to proceed even for large
domain deformation when the mesh becomes severely distorted. The mesh nodes are
constrained to the plane of the boundary they belong to for all but the free
surface.  

We adapt the same physical boundary conditions used above to the complete
(nonlinear) 2D problem. Thus, we write the kinematic condition as:
\begin{equation}
	\left( {\bf v} - \frac{\partial h}{\partial t} \right) \cdot \bf n = 0,
\end{equation}
$\bf n$ being the external unit normal vector. Both surface tension and
disjoining pressure exert normal stresses at the liquid-air interface
\begin{equation}
	{\cal S}\cdot {\bf n} = \left( \sigma {\cal C} - \kappa f(h) \right) {\bf n},
	\label{eq:pre_sup}
\end{equation}
where ${\cal C} = -{\nabla}_s \cdot {\bf n}$ is the curvature of the free
surface, $\nabla_s = \bf I_s \cdot \nabla$ is the surface gradient operator, and
${\bf I}_s= {\bf I} - {\bf n \, n}$ the surface identity tensor. At the ends of
the domain ($x = 0$ and $x = d$) periodic boundary conditions are applied for
both the velocity field and shape of the free surface. On the liquid--solid
interface, the the no-slip and no-penetration conditions ($\bf v = 0$) are
applied.

Since we must have the same length scale in both $x$ and $z$ directions in the
solution of the full non-linear N--S equations, we define now a slightly
different dimensionless set of units than in LSA for the long wave approximation
(see Eq.~(\ref{eq:scales1D})). Thus, the dimensionless variables in the
numerical simulations are given by:
\begin{equation}
	z=\ell \tilde z, \quad x=\ell \tilde x, \quad t=\tau \tilde t ,  \quad
	u=\frac {\gamma}{\mu} \tilde u, \quad  w=\frac {\gamma}{\mu} \tilde w,  
	\quad p= \frac {\gamma}{\ell} \tilde p, 
	\label{eq:adim1}
\end{equation}
which yields the dimensionless form of Navier--Stokes equations
\begin{eqnarray}
	La \left( \frac{\partial \tilde u}{\partial \tilde t} + 
	\tilde u \frac{\partial \tilde u}{\partial \tilde x} +
	w\frac{\partial \tilde u}{\partial \tilde z} \right) &=&- 
	\frac{\partial \tilde p}{\partial \tilde x} + 
	\left( \frac{\partial^2 \tilde u}{\partial {\tilde x}^2}+
	\frac{\partial^2 \tilde u}{\partial {\tilde z}^2} \right) ,
	\label{eq:NSx}\\
	La  \left( \frac{\partial \tilde w}{\partial \tilde t} + 
	\tilde u\frac{\partial \tilde w}{\partial \tilde x} +
	\tilde w\frac{\partial \tilde w}{\partial \tilde z} \right)  &=&- 
	\frac{\partial \tilde p}{\partial \tilde z} + 
	\left( \frac{\partial^2 \tilde w}{\partial {\tilde x}^2}+
	\frac{\partial^2 \tilde w}{\partial {\tilde z}^2} \right).
	\label{eq:NSz}
\end{eqnarray}
In particular, the boundary condition at the free surface,
Eq.~(\ref{eq:pre_sup}), becomes
\begin{equation}
	\tilde {\cal S}\cdot  {\bf n} = \left( \tilde {\cal C} - \frac{\varepsilon}{g_0} f(\tilde h) \right) {\bf n}.
	\label{eq:pre_sup1}
\end{equation}

\begin{figure*}
	\centering
	\includegraphics[width=.6\textwidth]{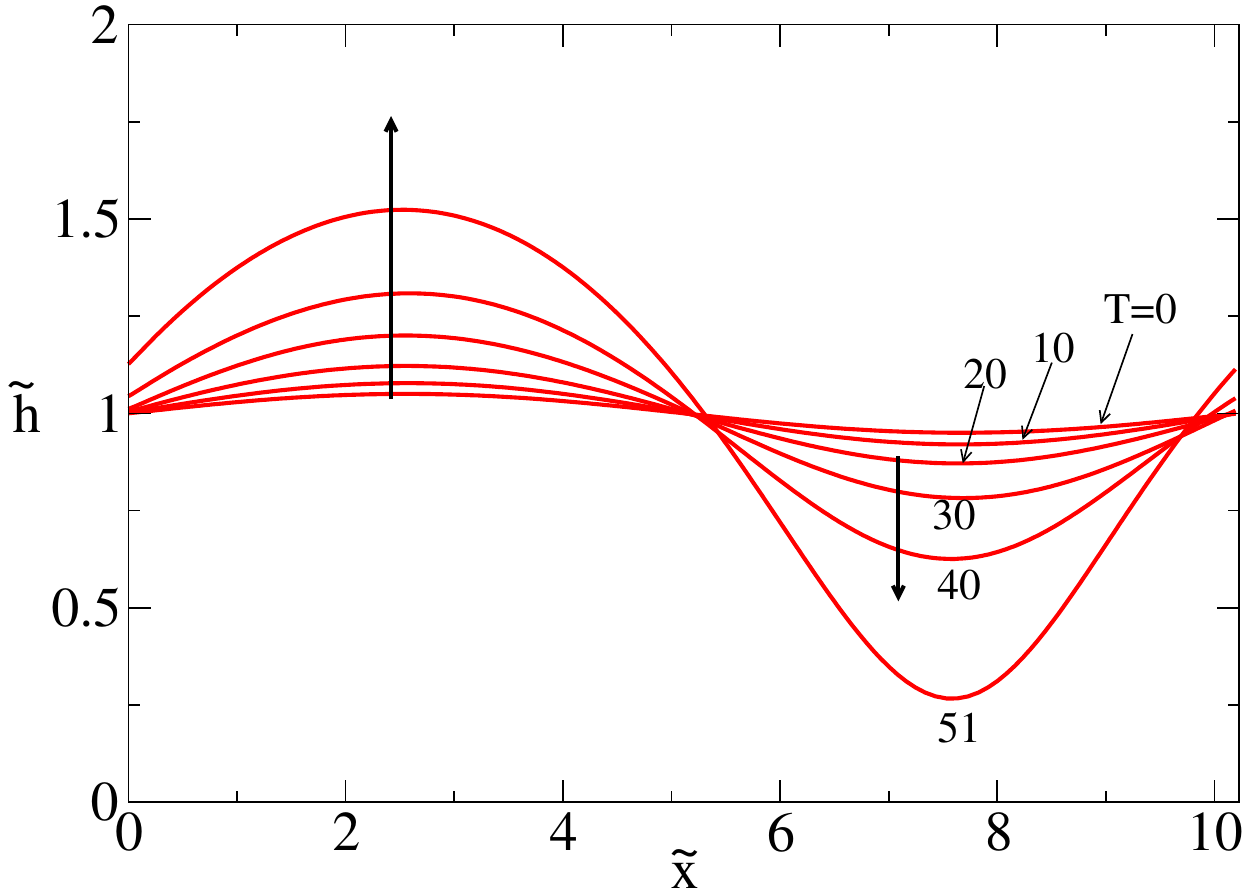}
	\caption{
		Time evolution of thickness profile for $La=1$ and $\varepsilon=1$. We use
		$A_0=0.05$.}
	\label{fig:hxLa1e1}
\end{figure*}

In order to observe the evolution of the mode with maximum growth rate of the 2D
model, we choose the length of the domain size in $x$-direction as 
$\tilde d =\tilde \lambda_m = 2 \pi / K_m^{2D}$. Note that this value is not
coincident with $K_m^{1D}$ (see Fig.~\ref{fig:km}). Thus, we use the following
monochromatic initial perturbation of the free surface 
\begin{equation}
	\tilde h(x,t=0)= \varepsilon + A_0 \sin \left( \frac{2 \pi \tilde x}{\tilde d} \right),
\end{equation}
where $A_0$ is a small amplitude ($A_0=0.05$ in the present calculations). In
Fig.~\ref{fig:hxLa1e1} we show a time evolution of the thickness profile for
$La=1$ and $\varepsilon=1$ (we use $(n,m)=(3,2)$ and $\tilde h_{\ast}=10^{-2}$
in all the following cases, unless otherwise stated). We carry on the simulation
until the film becomes too close to $\tilde h_{\ast}$, where the numerical
method is unable to converge, although continuation is sometimes possible by
using automatic remeshing.

\begin{figure*}
	\centering
	\includegraphics[width=\textwidth]{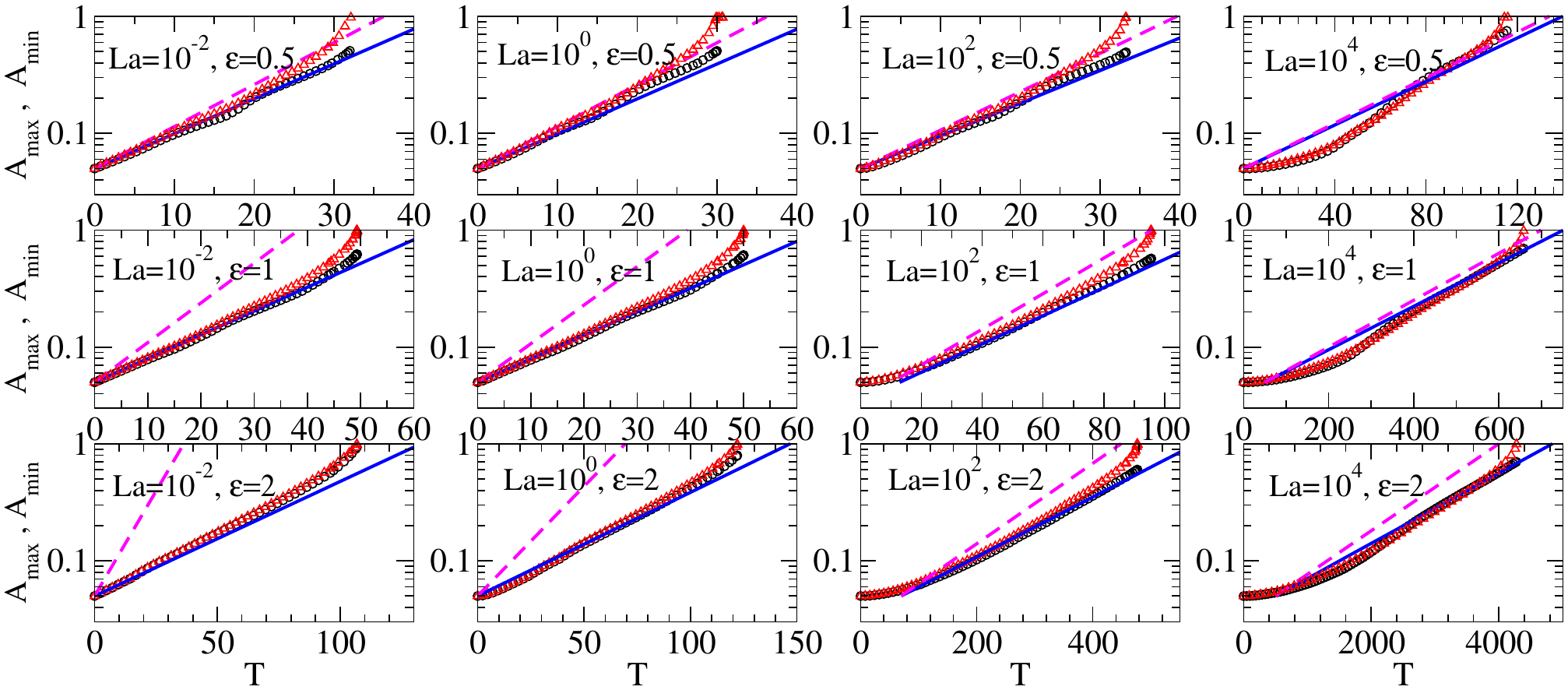}
	\caption{Time lines of the amplitudes $A_{max}$ ($\circ$) and $A_{min}$ 
		(\textcolor{red}{$\vartriangle$}) with $T={\tilde t} \varepsilon^3$ for
		different values of $La$ and $\varepsilon$. The lines correspond to the
		exponential behavior $A=0.05 \, \exp[\Omega_m (T-T_0)]$, where $\Omega_m$
		corresponds to the value given by either the 2D (solid line) or 1D (dashed line)
		model. $T_0 = 0$ except for the cases $(La,\varepsilon)=(10^4,0.5)$, $(10^2,1)$
		and $(10^4,1)$.}
	\label{fig:ampl_seq}
\end{figure*}

We study the evolution of the instability by tracking the maximum and minimum
amplitudes of the free surface deformation by defining
\begin{eqnarray}
	A_{max}(\tilde t)&=& \max_{0\leq \tilde x \leq \tilde L} \left|1 - \frac {\tilde h(\tilde t)}{\epsilon} \right|, \\ 
	A_{min}(\tilde t)&=& \min_{0\leq \tilde x \leq \tilde L} \left|1 - \frac {\tilde h(\tilde t)}{\epsilon} \right|. 
\end{eqnarray}
These results are plotted in Fig.~\ref{fig:ampl_seq} for the same values of $La$
used in Fig.~\ref{fig:RD_seq}, but $\varepsilon =0.5, 1, 2$. The numerical
non-linear solution of the problem shows that both $A_{max}$ and $A_{min}$ are
practically coincident during a relatively long time of the evolution. Within
the wide ranges of $La$ and $\varepsilon$ shown in Fig.~\ref{fig:ampl_seq}, this
behavior is observed for at least two thirds of the total time required for the
full development of the instability. This indicates that a linear models, like
those presented previously, are relevant to describe the flow beyond the onset
of the instability.

In order to compare the numerical results with the linear models, we plot in
Fig.~\ref{fig:ampl_seq} the expected exponential behavior as,
\begin{equation}
	A=0.05 \, e^{\alpha  (T-T_0)}
	\label{eq:expon}
\end{equation}
where $\alpha$ is given by the predicted growth rate for $K_m^{2D}$, and $T_0$
stands for a possible time shifting. For 1D model, $\alpha$ is the corresponding
growth rate for $K_m^{2D}$, i.e. $\alpha=\Omega^{1D}(K_m^{2D})$, which in
general does not coincide with the maximum growth rate within this
approximation. For 2D model, $\alpha=\Omega^{2D}(K_m^{2D})=\Omega_m$, which is
indeed the absolute maximum for this approach. Moreover, after separation of
$A_{max}$ and $A_{min}$, we expect that $A_{max}$ remains closer to the
exponential growth than $A_{min}$, which is more strongly affected by the
presence of the substrate. This effect is certainly observed in the numerical
results.

Figure~\ref{fig:ampl_seq} shows that for small $La$, say $La=0.01$ and $1$,
there is a very good agreement with the exponential behavior of the 2D model
prediction (solid blue lines). In general, the 1D model is not a good
approximation, except for very small $\varepsilon$, as expected. For both
models, we use $T_0=0$ since the behavior of $A_{max}$ and $A_{min}$ is of the
exponential type from the very beginning. This type of growth is also observed
for $La=100$ and $\varepsilon=0.5$, but $T_0 \neq 0$ is needed for large
$\varepsilon$, thus indicating the presence of a very early stage with slower
(non exponential) growth. This effect is still more pronounced for $La$ as large
as $La=10^4$. In these cases, where there is still an acceptable agreement
between the 2D model and the numerics for relatively large $\varepsilon$.
However, for this very large value of $La$, as $\varepsilon$ is decreased
neither 1D nor 2D models are able to capture the actual evolution of the
complete nonlinear problem. This issue deserves further investigation, which is
out of the scope of the present paper and remains for future work.

\begin{figure}
	\centering
	\subfigure[] {\includegraphics[width=.47\textwidth]{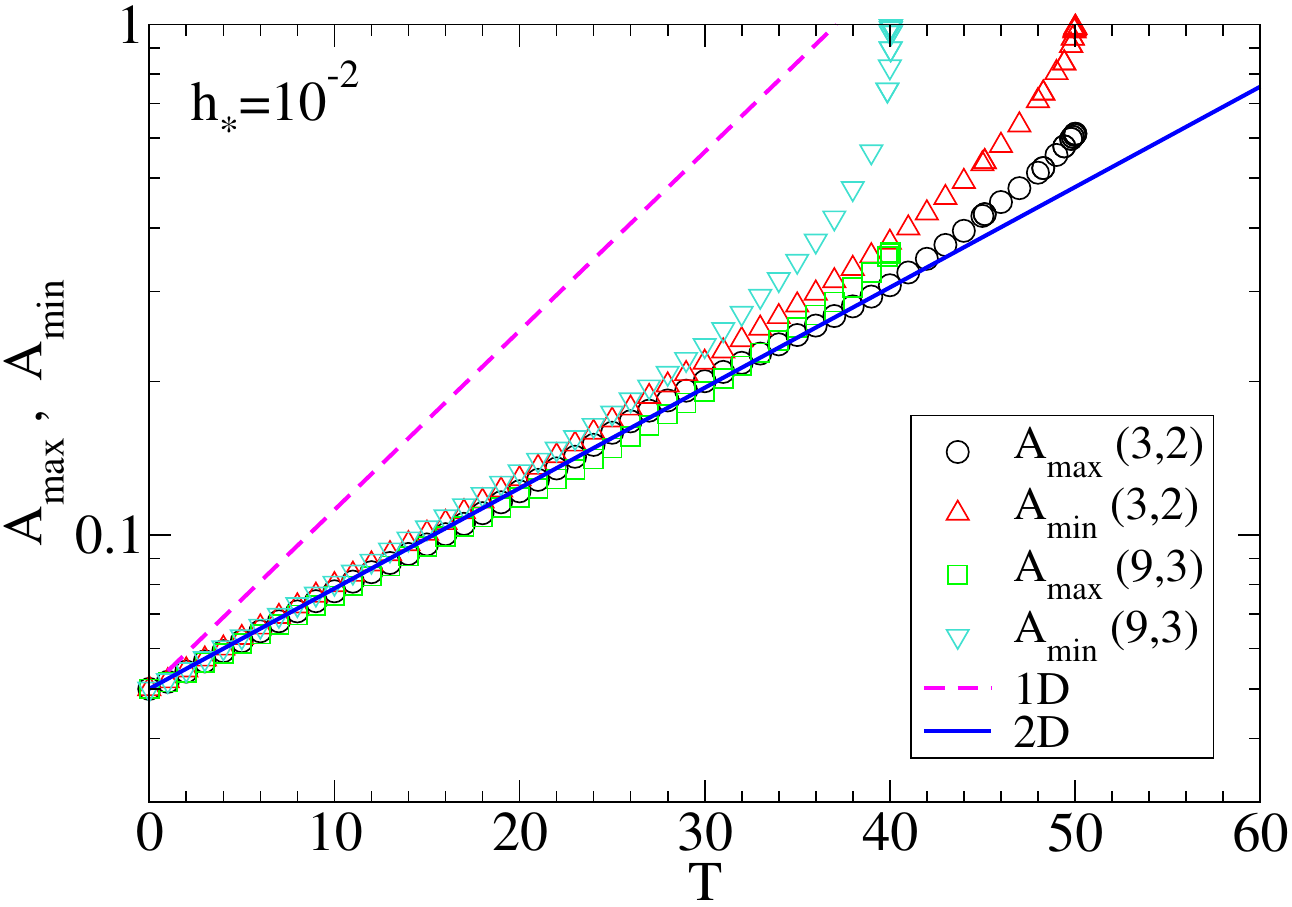}}
	\subfigure[] {\includegraphics[width=.47\textwidth]{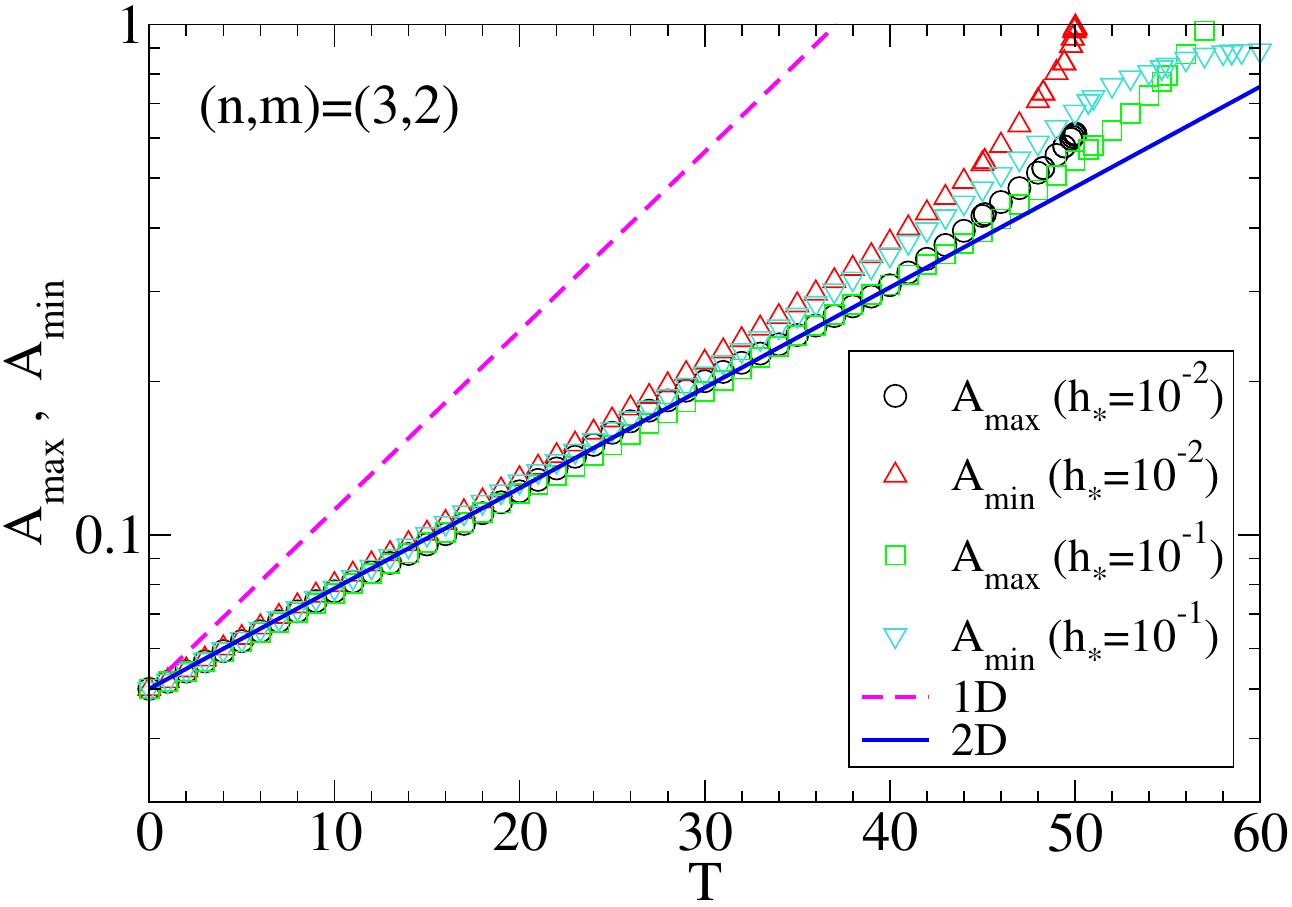}}
	\caption{Time lines of the amplitudes $A_{max}$ and $A_{min}$ for $La=1$ and
		$\varepsilon =1$ for: (a) two different pairs of exponents $(n,m)$, (b) two
		different values of $\tilde h_\ast$. Symbols indicate numerical simulations,
		lines the predictions of linear models.}
	\label{fig:La1_e1}
\end{figure}

It is worth noting that the exponents $n$ and $m$ of the disjoining pressure in
Eq.~(\ref{eq:Pih}) do not a play a role in both linear analyses performed here.
Their influence in this stage is somehow hidden in the length scale, $\ell$,
defined in Eq.~(\ref{eq:ell}). However, some effects are expected in the
numerical solution of the fully nonlinear N--S equations, since they appear in
the boundary condition given by Eq.~(\ref{eq:pre_sup1}).
Figure~\ref{fig:La1_e1}a shows a comparison of the time evolution of $A_{max}$
and $A_{min}$ for $(n,m)=(3,2)$ and $(9,3)$, which are typical pairs of
exponents used in the literature~\citep{schwartz_lang98}. Clearly, both cases are practically coincident
in the linear stage, and are in agreement with the linear 2D model. For larger
times, the corresponding non-linear regimes strongly differ, thus leading to
different breakup times, so that the effect of the exponents is limited to the
short final non-linear stage. Similarly, Fig.~\ref{fig:La1_e1}b shows the same
time lines for two different values of $\tilde h_{\ast}$ and a given pair of
$(n,m)$. Also in this case, only the nonlinear stage of the evolution changes
for different thicknesses $\tilde h_{\ast}$, without any significant change of
the early linear stage. For $\tilde h_{\ast}$ as small as $\tilde
h_{\ast}=10^{-3}$, no difference is observed neither in the linear nor in the
nonlinear stages. This so because $\tilde h_{\ast}$ becomes negligible with respect
to $\tilde h_0$.

\section{Summary and conclusions}
In this work, we have developed three different approaches to study the
instability of a flat liquid thin film under partial wetting conditions, and
subject to intermolecular forces (disjoining pressure): long wave 1D model (with
inertia), liner 2D model, and fully nonlinear numerical simulations. Firstly, we
have extended the purely viscous analysis within the lubrication approximation
to one where inertial effects are taken into account, which we call for brevity
1D model. The LSA of this model shows that inertia does not lead to new regions
of instability compared with the purely viscous case. Instead, it adds new
stable modes: some which are exponentially decaying, and others which are damped
oscillations. The former extend over the same range of the unstable modes and
even beyond, while the latter appear for larger wavenumbers. In the unstable
region of most interest here, we find that both the marginal wavenumber and that
of the maximum growth rate do not change at all with the addition of inertia.
However, the results clearly show that the growth rates of the instability
decrease as inertial effects are stronger. The intensity of these effects is
here quantified by a single parameter, namely the modified Laplace number,
$La^\ast=La \,\varepsilon^5$. Therefore, the approximation can be applied only
for large $La$, since $\varepsilon \ll 1$ is required for the approach to be
valid. 

Secondly, we develop a LSA of the Navier--Stokes equation, so that the
restriction of small aspect ratio, $\varepsilon$, is no longer required. This
calculation, called for brevity 2D model, is particularly useful to assess the
accuracy of the 1D model predictions. The main difference between these models
is the way that inertia is treated. In linear 1D model, the convective terms for
the horizontal direction are still taken into account, while horizontal and
vertical convective terms are neglected in the linear 2D model, but the viscous
Laplacian term is now fully conserved for both directions. Thus, we have now two
independent parameters to characterize the flow, namely $La$ and $\varepsilon$.
The 2D model shows that the marginal wavenumber remains the same as 1D model,
and does not depend on $La$. However, unlike 1D model, 2D model shows that $K_m$
is not constant, and decreases as $La$ increases. This is an important result,
since it shows that inertia can modify the distance between the final drops,
which must be more separated with respect to the purely viscous case.

With respect to the dependence of the growth rates with $La$, 2D model also
shows that they decrease for increasing $La$, but the strength of the effect is
greater than what is predicted by 1D model. Interestingly, the discrepancies between
both models decrease as $La$ increases, i.e. for larger inertial effects. Note
also that both models capture the main scaling of the dimensional growth rate,
$\omega$, with the aspect ratio $\varepsilon$. Thus, we can write, 
\begin{equation}
	\omega_1(k) = \frac{\varepsilon^3}{\tau} \Omega_1(k \ell;La \, \varepsilon^5), \quad
	\omega_2(k) = \frac{\varepsilon^3}{\tau}  \Omega_2(k \ell ;La, \varepsilon)
\end{equation}
where the subscripts $1$ and $2$ correspond to 1D and 2D models, respectively.

\begin{figure}
	\centering
	\includegraphics[width=.45\textwidth]{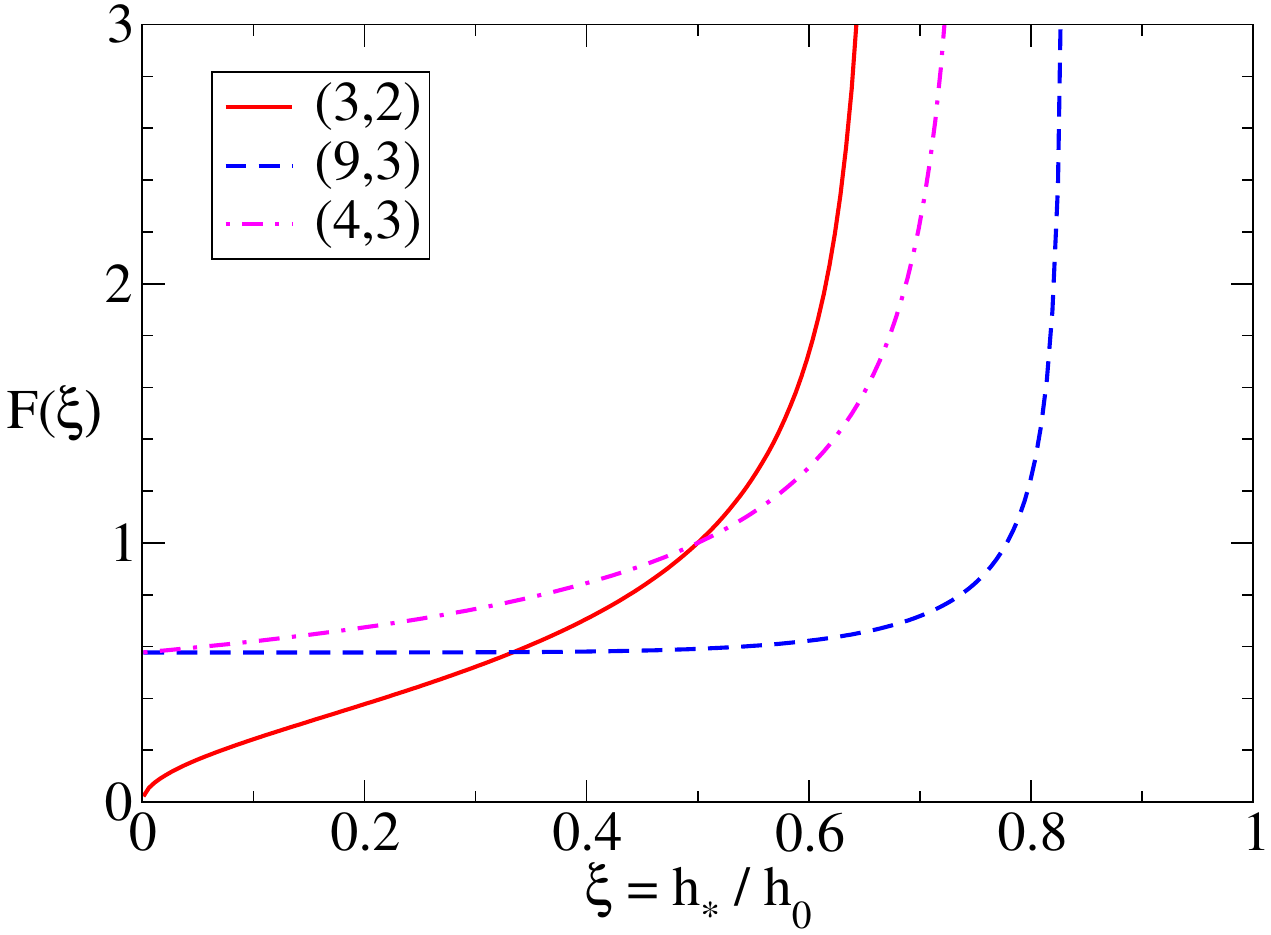}
	\caption{Function $F(\xi)$ that determines the influence of $h_{\ast}$ on the
		characteristic length scale, $\ell$.}
	\label{fig:F}
\end{figure}

Finally, we are concerned now with obtaining a prediction of the both $k_m$ and
$\omega_m$ as a function of the film thickness, $h_0$, for a given experimental
configuration. In order to do so, we recall that~\citep{Israelachvili} 
\begin{equation}
	\kappa = \frac {A}{6 \pi h_{\ast}^3},
	\label{eq:kA}
\end{equation}
where $A$ is the Hamaker constant. Thus, the characteristic length, $\ell$,
given by Eq.~(\ref{eq:ell}) can be written as
\begin{equation}
	\ell = F(\xi) \frac{h_0^2}{L},
\end{equation}
where
\begin{equation}
	L=\sqrt{ \frac{A}{6 \pi \gamma} }, \qquad
	F(\xi)= \sqrt{ \frac{\xi^3}{g_0(\xi)}},
\end{equation}
being $\xi= h_{\ast}/h_0$. The function $F(\xi)$, which describes the effects of
$h_{\ast}$, is shown in Fig.~\ref{fig:F} for three usual values of $(n,m)$.
Interestingly, the cases with $m=3$ and large $n$ present a practically constant
region for $\xi <0.5$, which is a typical range in experiments. In these cases,
we notice that $\ell \propto h_0^2$, thus $\varepsilon \propto h_0^{-1}$ and $La
\propto h_0^2$. Instead, if $m<3$, say $m=2$, the behaviour is different since
$F \rightarrow 0$ for decreasing $h_{\ast}$. For $(3,2)$, we have $\ell \propto
h_0^{3/2}$, thus $\varepsilon \propto h_0^{-1/2}$ and $La \propto h_0^{3/2}$. 

\begin{figure}
	\centering
	\subfigure[]{\includegraphics[width=.46\textwidth]{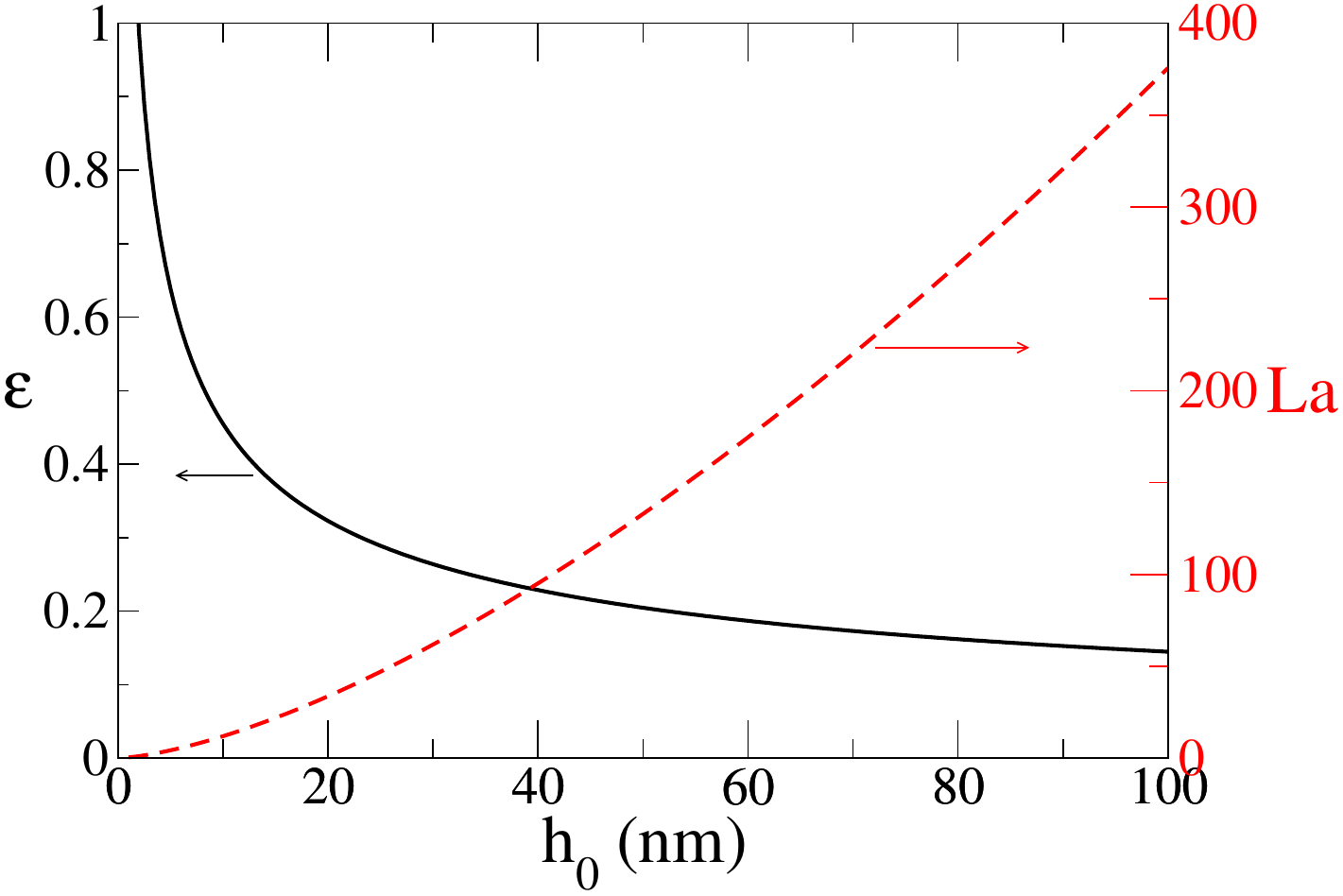}}
	\subfigure[]{\includegraphics[width=.46\textwidth]{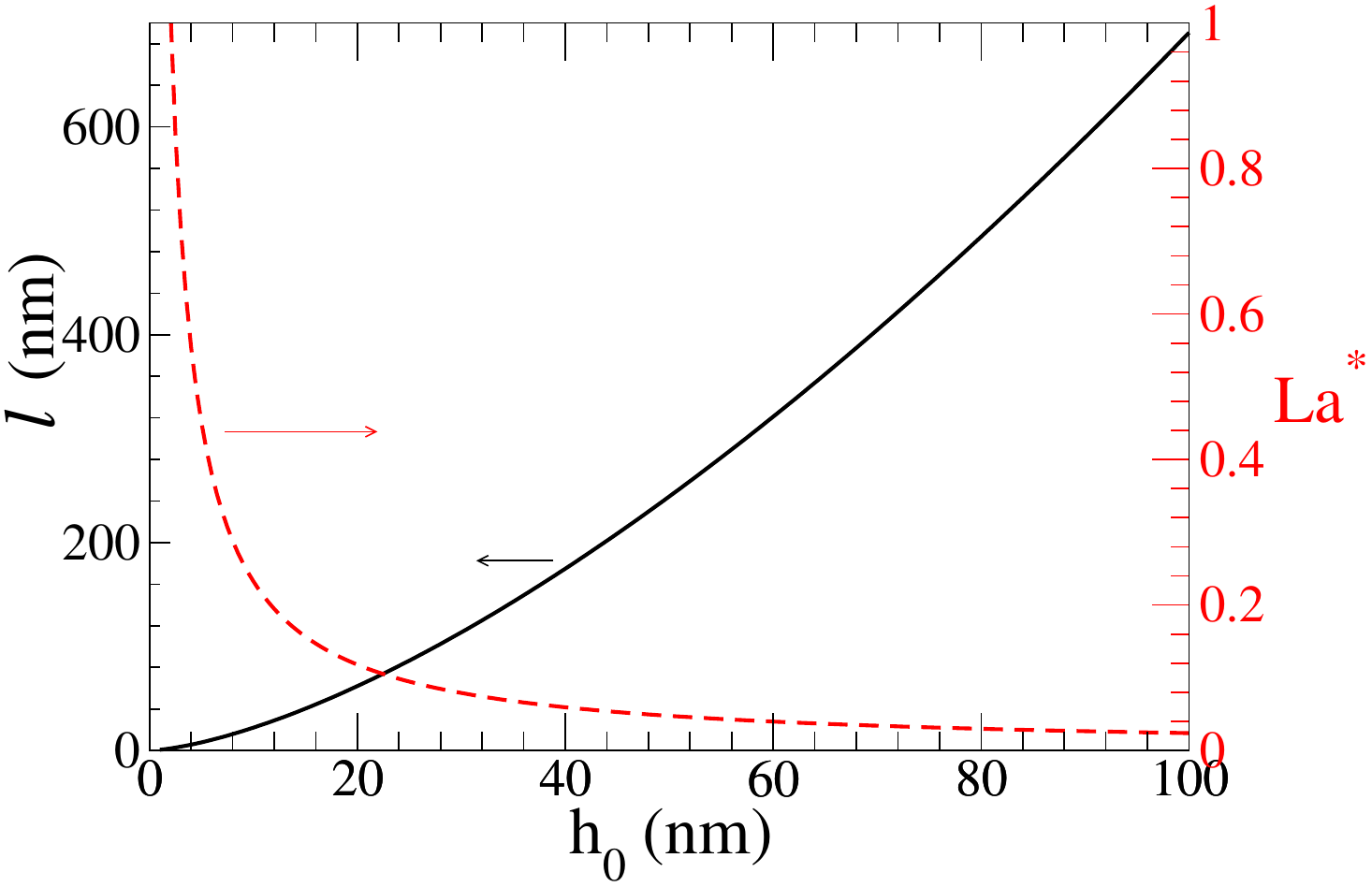}}
	\caption{Dependence of the dimensionless parameters $\varepsilon$, $La$,
		$La^\ast$, and the characteristic length scale, $\ell$, as a function of the
		film thickness, $h_0$, for melted copper films}.
	\label{fig:param}
\end{figure}

These results shoud be taken into account when analyzing experimental data
within a given hydrodynamic model. For instance, the lubrication approximation
would not become more valid as $h_0$ decreases (as it could be expected {\it a
priori}) since $\varepsilon$ increases for thinner films. In fact, let us
consider the data from the experiments with melted copper films on a $SiO_2$
substrate reported in~\citep{gonzalez_lang2013}. In this case, we have
$\gamma=1.304\,Kg/m^3$, $\mu=0.00438\,Pa\,s$, and the experiments could be
fitted with a purely viscous lubrication model using $A=2.58~10^{-19}\,J$,
$h_\ast=0.1\,nm$ and $(n,m)=(3,2)$. Thus, we calculate the corresponding values
of $\varepsilon$ and $La$ for film thickness, $h_0$, in the interval
$(1,100)\,nm$, as shown Fig.~\ref{fig:param}a. Note that even if inertial
effects increase as $h_0$ increases, $\varepsilon$ decreases even faster, so
that lubrication approximation assumptions apply for larger $h_0$'s (see also
$La^\ast$ in Fig.~\ref{fig:param}b). Consistently, Fig.~\ref{fig:param}b
indicates that the length $\ell$ (proportional to the critical wavelength)
increases with $h_0$, so that wavelengths of some hundreds of nanometers should
be expected for these film thicknesses. 
\begin{figure}[htb]
	\centering
	\includegraphics[width=.6\textwidth]{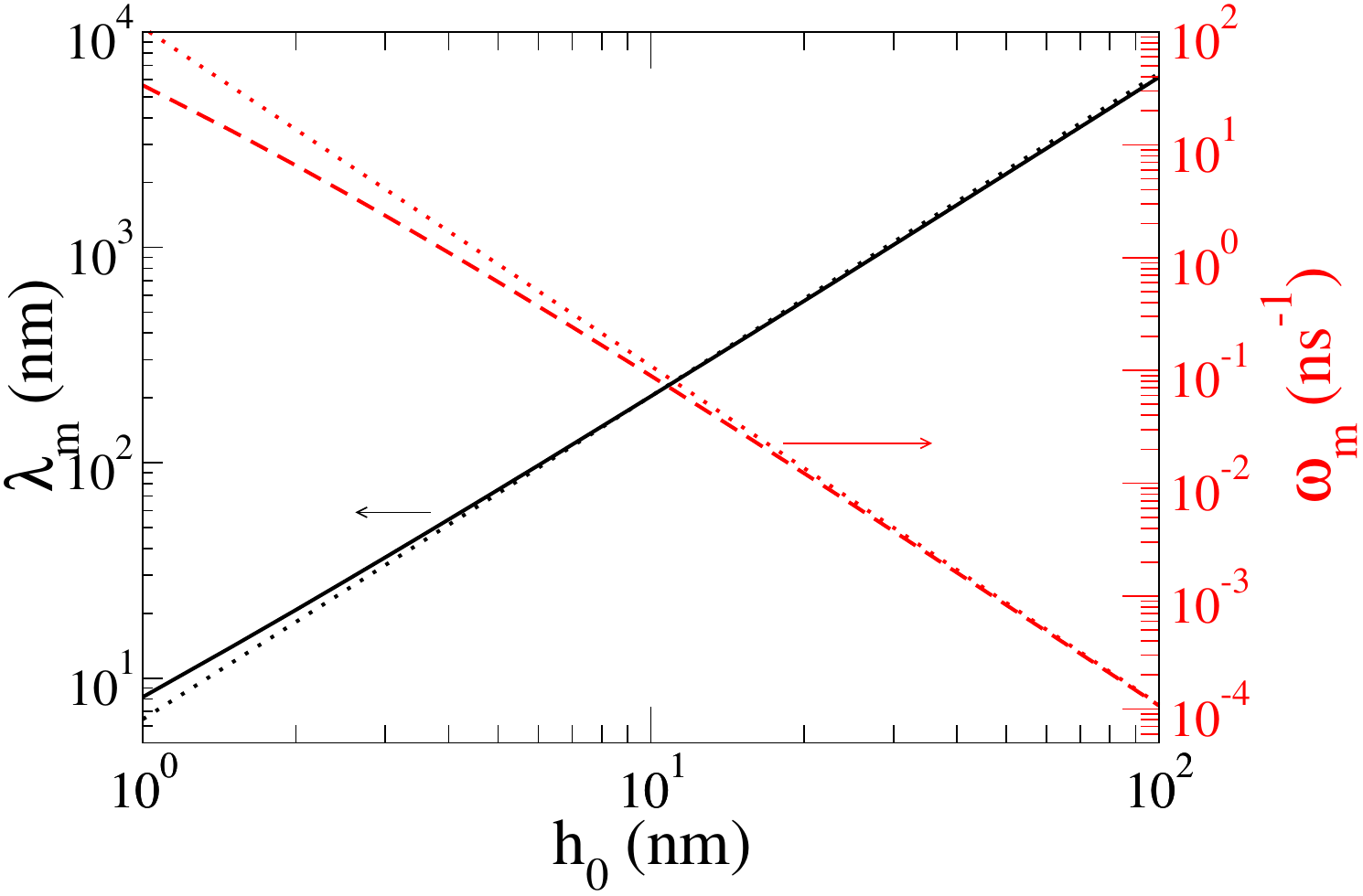} 
	\caption{Wavelength of maximum growth rate, $\lambda_m$, and the corresponding
		growth, $\omega_m$, as a function of the film thickness, $h_0$, for melted
		copper films}
	\label{fig:lm-wm}
\end{figure}

In particular, we show the wavelength of maximum growth rate, $\lambda_m$, as
well as the corresponding growth, $\omega_m$, as a function of $h_0$ in
Fig.~\ref{fig:lm-wm}. The asymptotic power laws for large $h_0$, given by the
lubrication approximation where $K_m=1/\sqrt{2}$ and $\Omega_m=1/12$ are (see
also Eq.~(\ref{eq:scales1D})),
\begin{equation}
	\lambda_m = 2 \pi \sqrt{\frac{L^2}{h_\ast}} h_0^{3/2}, \qquad
	\omega_m = \frac{\gamma}{3 \mu} \frac{L^4}{h_{\ast}^2} h_0^{-3}.
\end{equation}
These expressions are plotted as dotted lines in Fig.~\ref{fig:lm-wm}. Therefore,
we conclude that both inertial and bidimensional effects are not significant
if $h_0 \gtrsim 20 h_{\ast}$, being pretty safe to use lubrication approximation
results to describe the instability even for large $La$ provided $h_0 \gg
h_{\ast}$, as in the experiments reported by \cite{gonzalez_lang2013}. However,
for very thin nanometric films with $h_0 \lesssim 10 h_{\ast}$, these effects
should be taken into account, specially when analyzing the growth rates of the
unstable modes.

\section*{Acknowledgements}
A.G.G. and J.A.D. acknowledge support from Consejo Nacional de Investigaciones
Cien- t\'{\i}ficas y T\'ecnicas de la Rep\'ublica Argentina (CONICET, Argentina)
with grant PIP 844/2011 and Agencia Nacional de Promoci\'on de Cient\'{\i}fica y
Tecnol\'ogica (ANPCyT, Argentina) with grant PICT 931/2012. MS gratefully
acknowledges the College of Engineering at University of Canterbury for its
financial support to visit A.G.G. and J.A.D. in Argentina.  

\bibliographystyle{chicago}

\end{document}